\documentclass[fleqn,usenatbib]{mnras}

\usepackage{newtxtext,newtxmath}

\usepackage[T1]{fontenc}
\usepackage{ae,aecompl}

\usepackage{multirow}
\usepackage{subfig}

\usepackage{graphicx}	
\usepackage{amsmath}	
\usepackage{amssymb}	

\title[Filling and Abundance Discrepancy Factors]{Realistic Models for Filling and Abundance Discrepancy Factors in Photoionised Nebulae}

\author[Bergerud, Spangler, \& Beauchamp]{
Brandon M. Bergerud,$^{1}$\thanks{E-mail: brandon-bergerud@uiowa.edu}
Steven R. Spangler,$^{1}$
and Kara M. Beauchamp$^{2}$
\\
$^{1}$Department of Physics and Astronomy, The University of Iowa, Iowa City, IA 52242\\
$^{2}$Department of Physics and Engineering, Cornell College, Mount Vernon, IA 52314\\
}

\date{Accepted XXX. Received YYY; in original form ZZZ}

\pubyear{2019}

\begin{document}
\label{firstpage}
\pagerange{\pageref{firstpage}--\pageref{lastpage}}
\maketitle

\begin{abstract}
When comparing nebular electron densities derived from collisionally excited lines (CELs) to those estimated using the emission measure, significant discrepancies are common. The standard solution is to view nebulae as aggregates of dense regions of constant density in an otherwise empty void. This porosity is parametrized by a filling factor $f<1$. Similarly, abundance and temperature discrepancies between optical recombination lines (ORLs) and CELs are often explained by invoking a dual delta distribution of a dense, cool, metal-rich component immersed in a diffuse, warm, metal-poor plasma. In this paper, we examine the possibility that the observational diagnostics that lead to such discrepancies can be produced by a realistic distribution of density and temperature fluctuations, such as might arise in plasma turbulence. We produce simulated nebulae with density and temperature fluctuations described by various probability distribution functions (pdfs). Standard astronomical diagnostics are applied to these simulated observations to derive estimates of nebular densities, temperatures, and abundances. Our results show that for plausible density pdfs the simulated observations lead to filling factors in the observed range. None of our simulations satisfactorily reproduce the abundance discrepancy factors (ADFs) in planetary nebulae, although there is possible consistency with \ion{H}{ii} regions. Compared to the case of density-only and temperature-only fluctuations, a positive correlation between density and temperature reduces the filling factor and ADF (from optical CELs), whereas a negative correlation increases both, eventually causing the filling factor to exceed unity. This result suggests that real observations can provide constraints on the thermodynamics of small scale fluctuations.
\end{abstract}




\section{Introduction}
An understanding of plasmas in photoionised regions such as planetary nebulae (PNe), \ion{H}{ii} regions, Wolf-Rayet nebulae, and Active Galactic Nuclei requires knowledge of fundamental plasma parameters, such as the electron density, electron temperature, ion temperature, and magnetic field strength. The electron density may be measured spectroscopically from the ratio of collisionally excited lines (CELs), or from the emission measure available from H$\alpha$, H$\beta$, or radio continuum brightness measurements, which is the integrated electron density squared through the nebula along the line of sight,
\begin{equation}\label{eq:emissionMeasure}
\mathrm{EM} = \int n_{e}^{2} \, ds.
\end{equation}
The spectroscopic density estimates are generally larger than those found using the emission measure \citep{1984ApJ...287..116K, 1988RMxAA..16..111M, 1994A&A...284..248B, 2003ApJ...587..562M, 2016MNRAS.460.4038E}. The common solution to this discrepancy is to invoke a `filling factor', which was first introduced by \cite{1959ApJ...129...26O}.\footnote{When Osterbrock \& Flather proposed a filling factor for the Orion Nebula $(f\approx 0.03)$, they did so under an assumption that is no longer tenable. More recent studies have indicated that the nebula is sheet-like rather than spherical \citep{1995ApJ...438..784W}, and when correcting for geometry the density discrepancy is greatly reduced.} In this model, nebulae are considered to be aggregates of dense regions of constant density in an otherwise empty void. The volume percentage that these collections make up is called the filling factor. It thus reconciles the two density estimates by restricting the emitting gas to only a fraction of the line of sight through the nebula. In practice, the filling factor is computed as the ratio of the squared density estimates,
\begin{equation}\label{eq:fillingFactor}
f = \frac{n_{e, \mathrm{EM}}^{2}}{n_{e, \mathrm{CEL}}^{2}}
\end{equation}
where $n_{e, \mathrm{CEL}}$ is the density derived from CELs and $n_{e, \mathrm{EM}}$ is the density derived from the emission measure,
\begin{equation}\label{eq:emissionMeasureDensity}
n_{e, \mathrm{EM}} \equiv \sqrt{\frac{\mathrm{EM}}{L}}
\end{equation}
where $L$ is the depth of the nebula along the line of sight. While there is a wide spread in observed values, $f = 0.3$ is commonly adopted for PNe \citep{1994A&A...284..248B, 2016MNRAS.455.1459F, 2017MNRAS.468.1794L}, while \ion{H}{ii} regions generally have smaller values \citep{1984ApJ...287..116K} whose filling factors are inversely related to the size of the nebula \citep{2013ApJ...765L..24C}.

This model, however, considerably oversimplifies the physical conditions of nebulae and may hinder an understanding of important dynamics in such regions. In fact, abundances (and temperatures) derived from optical recombination lines (ORLs) and CELs are another area of discrepancies that constitute the \emph{abundance discrepancy problem} \citep[e.g.][]{2000MNRAS.312..585L, 2006IAUS..234..219L, 2006MNRAS.368.1959L, 2007ApJ...670..457G, 2015MNRAS.453.1281B}. The abundance discrepancy is quantified in terms of the \emph{abundance discrepancy factor} (ADF), which is defined as the ratio of the ionic abundance derived from ORLs to the abundance derived from CELs \citep{2017PASP..129h2001P}:
\begin{equation}
\mathrm{ADF}\left(X^{+i}\right) \equiv \frac{X^{+i}_\mathrm{ORLs}}{X^{+i}_\mathrm{CELs}}.
\end{equation}
While the distribution of ADFs among \ion{H}{ii} regions is normally distributed $(\mu\approx 2, \;\sigma \approx 0.5)$, the distribution among PNe has a prominent tail extending to ADFs of nearly three orders of magnitude, with the more extreme cases frequently being associated with binary systems. When modelling the PNe distribution as consisting of a mixture of a normal distribution and an extreme distribution, the normal component has a similar mean and standard deviation as observed for \ion{H}{ii} regions, which could indicate a similar origin for the discrepancy \citep{2018MNRAS.480.4589W}. A common explanation for these discrepancies is to invoke a simplified model consisting of a dense, cool, metal-rich component immersed in a diffuse, warm, metal-poor plasma -- particularly for the more anomalous PNe observations \citep[e.g.][]{2000MNRAS.312..585L}.

Diagnostic measures inherently assume a constant density and temperature, while stochastic fluctuations in density, temperature, flow velocity, magnetic field, and other variables naturally arise in turbulent media such as the solar wind and the interstellar medium. In this paper, we explore the possibility that similar turbulence in photoionised nebulae could be responsible for the observed discrepancies between density, temperature, and abundance when using different diagnostics. In \S\ref{s:model}, we provide an overview of our model and discuss the various probability distribution functions (pdfs) and diagnostics that are used; \S\ref{s:simulations} examines the effects of density and temperature fluctuations in isolation, and concludes with the case when there is a polytropic relationship between the two. In \S\ref{s:conclusion} we conclude with a summary of our results. 

\section{Model}\label{s:model}

\subsection{Nebular Geometry}
We constructed a model nebula using a 3D grid of equally sized cells. For cells located within the nebula, densities and temperatures are assigned based on the chosen distribution function. These conditions are used to generate emission coefficients for each cell, which are then integrated along the line of sight to generate observed line intensities.

\begin{figure}
\includegraphics[width=0.99\columnwidth]{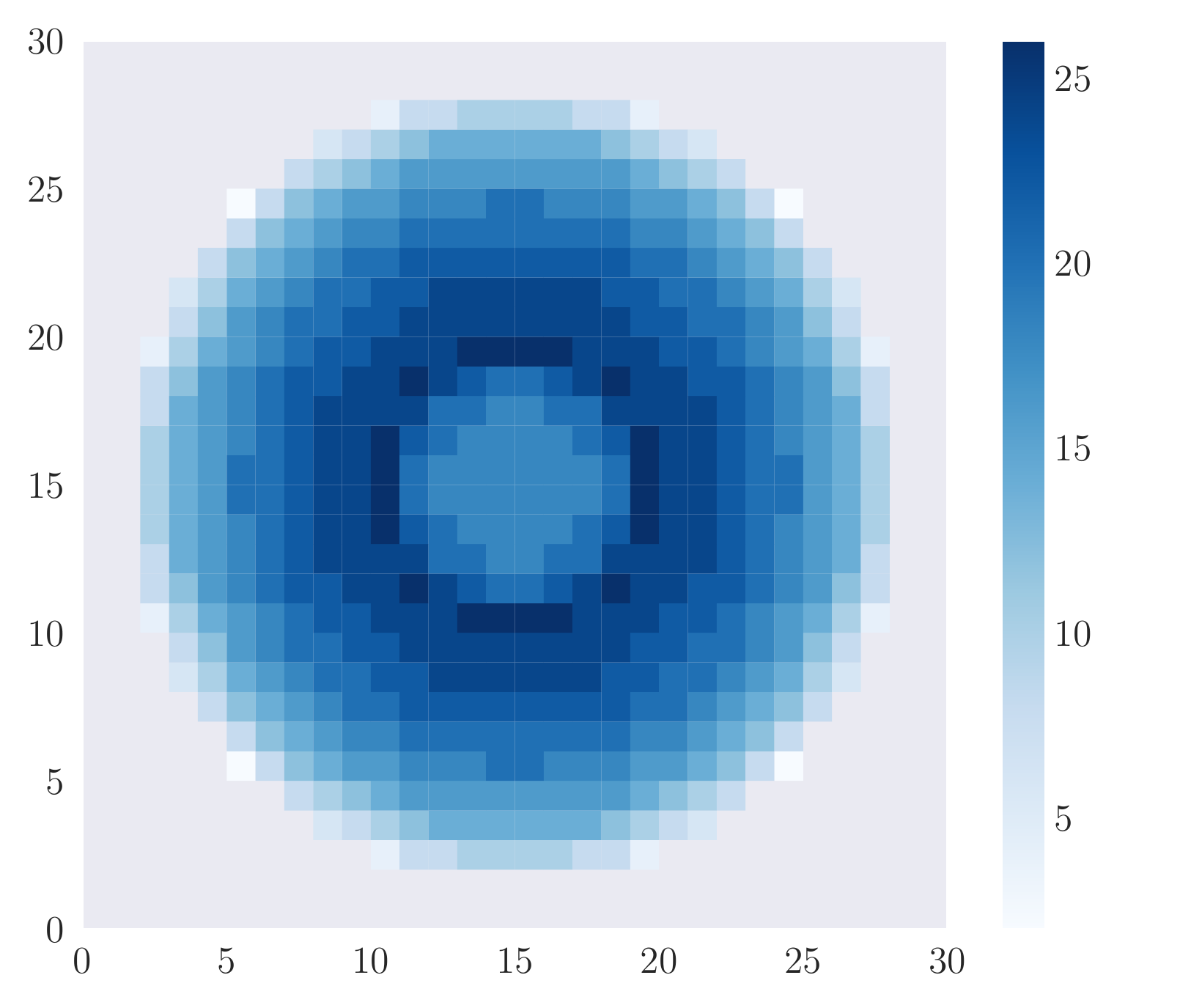}
\caption{A simulated spherical nebula containing a central cavity with a radius $\frac{1}{3}$ of the outer radius. The number of cells along each line of sight have been plotted.}
\label{fig:geomSphere}
\end{figure}

To define the extent of the nebula, a spherical model is used, parametrized by an inner and outer radius. Central cavities are commonly observed in nebulae, which are thought to result from strong stellar winds interacting with the surrounding medium \citep{1975ApJ...200L.107C, 1977ApJ...218..377W}. To simulate these cavities, we set the inner radius to be one-third of the outer radius based on the Rosette Nebula \citep{1985A&A...144..171C}, although the relative size of the central cavity was found to have little effect on our results. The line of sight depth for a simulated nebula is shown in Fig. \ref{fig:geomSphere}.

\subsection{Density Fluctuations}
Several pdfs are used to model inhomogeneous electron densities: exponential, lognormal, and power lognormal. The exponential distribution is amenable to analytical solutions, while lognormal densities are observed in the solar wind plasma \citep{2000JGR...105.2357B} and in turbulent molecular cloud simulations \citep{1997ApJ...474..730P, 2001ApJ...546..980O}. The power lognormal distribution is a modified version of the lognormal distribution, providing a more extended tail to the distribution.\footnote{Normal distributions naturally arise when an outcome $X$ depends on the sum of $N$ random variables $x_{i}$,  $X = \sum_{i=1}^{N} x_{i}$. If an outcome depends on their product, $X = \prod_{i=1}^{N} x_{i}$, then the lognormal distribution is a natural result since $\ln{X} = \sum_{i=1}^{N} \ln{x_i}$. Powerlaw distributions occur when there is a multiplicative growth subject to a random cessation event \citep{2012msma.book.....F}.}

For the exponential distribution, the probability a cell has an electron density $n_{e}=x$ is given by
\begin{equation}
p(x|\bar{x}) = \frac{e^{-x / \bar{x}}}{\bar{x}}
\end{equation}
where $\bar{x}$ is the mean value.

The lognormal distribution is an exponential transformation of a Gaussian distribution. If $\mu$ is the mean and $\sigma$ the standard deviation of a Gaussian pdf, the lognormal pdf is given by
\begin{equation}\label{eq:pdf_log}
p(x|\mu, \, \sigma) = \frac{1}{x \sqrt{2\pi\sigma^2}} \exp\left[-\frac{(\ln x - \mu)^2}{2\sigma^2}\right].
\end{equation}
While the median scales with the exponential, $\tilde{x} = e^{\mu}$, the mode shifts towards lower values, $\hat{x} = e^{\mu - \sigma^{2}}$, and the mean towards higher values, $\bar{x} = e^{\mu + \sigma^{2} / 2}$.

Pareto (powerlaw) distributions are ubiquitous in nature, having been applied to model diverse phenomena such as the stellar initial mass function \citep{2015MNRAS.449.2413B}, the size of rain shower cells
\citep{Machado2013}, and socio-economic factors such as income and population size \cite[e.g.][and references therein]{doi:10.1081/STA-120037438}. Powerlaw distributions often break down over part of the domain, and one such modification is the power lognormal distribution, whose pdf is given by\footnote{\url{https://docs.scipy.org/doc/scipy/reference/tutorial/stats/continuous_powerlognorm.html}}
\begin{equation}
p(x|\sigma, c) = \frac{c}{x\sigma}
\phi\left(\frac{\ln x}{\sigma}\right)
\left[\Phi\left(-\frac{\ln x}{\sigma}\right)\right]^{c-1}
\end{equation}
where $\phi(z)$ is the normal distribution and $\Phi(z)=\frac{1}{2}\mathrm{erfc}(-z)$ is the normal cumulative distribution function, where $\mathrm{erfc}$ is the complementary error function. After sampling, the densities are multiplied by a scale factor to achieve the desired mean density.\footnote{An alternative formulation is the modified lognormal power-law (MLP) distribution, which has been applied to the stellar initial mass function \citep{2015MNRAS.449.2413B}. The profile assumes an initial lognormal distribution, followed by an exponential growth with an exponential distribution of growth lifetimes.}

\begin{figure}
    \centering
    \subfloat[Lognormal]{\includegraphics[width=\columnwidth]{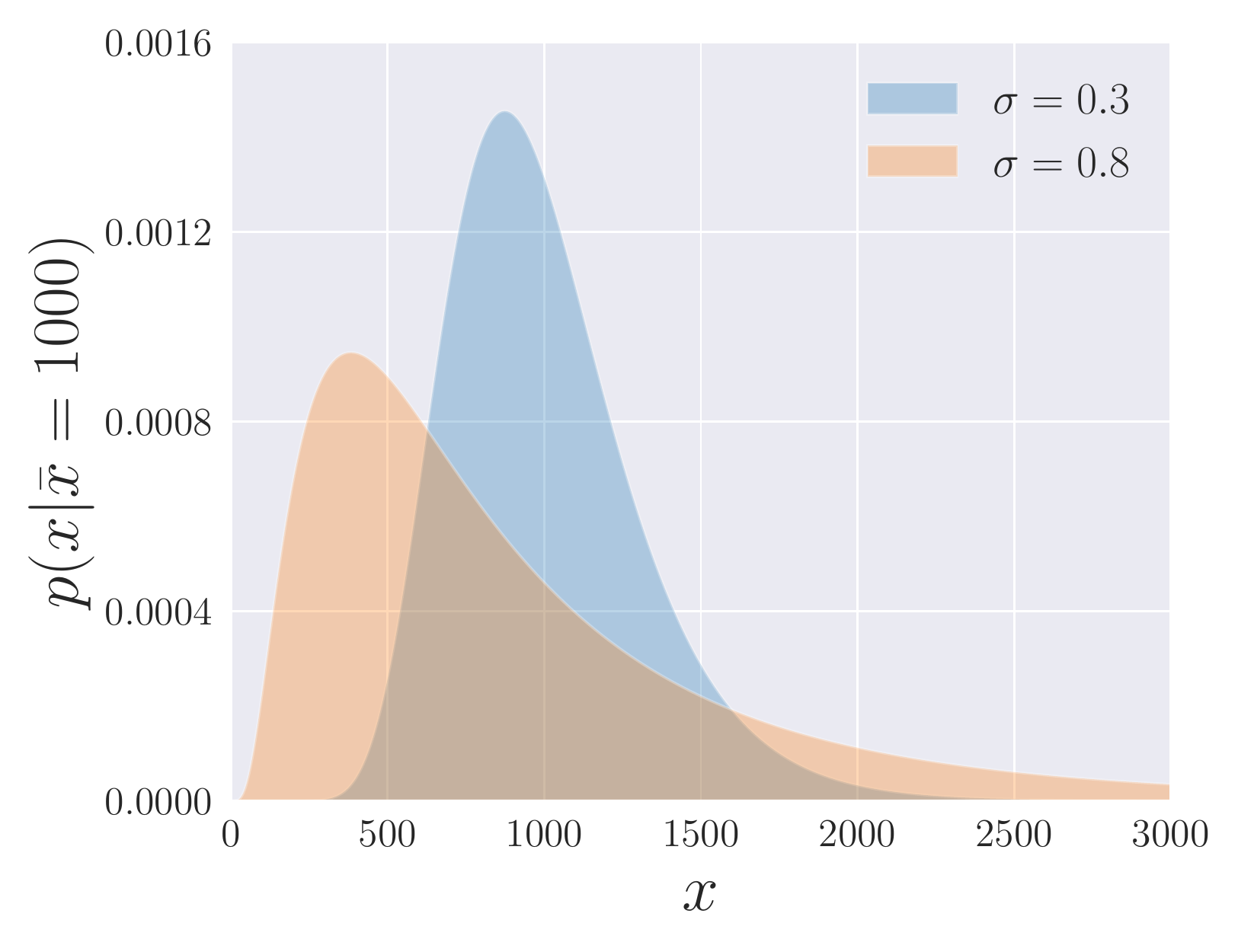} \label{fig:pdfs1}}
    \\
    \subfloat[Power Lognormal]{\includegraphics[width=\columnwidth]{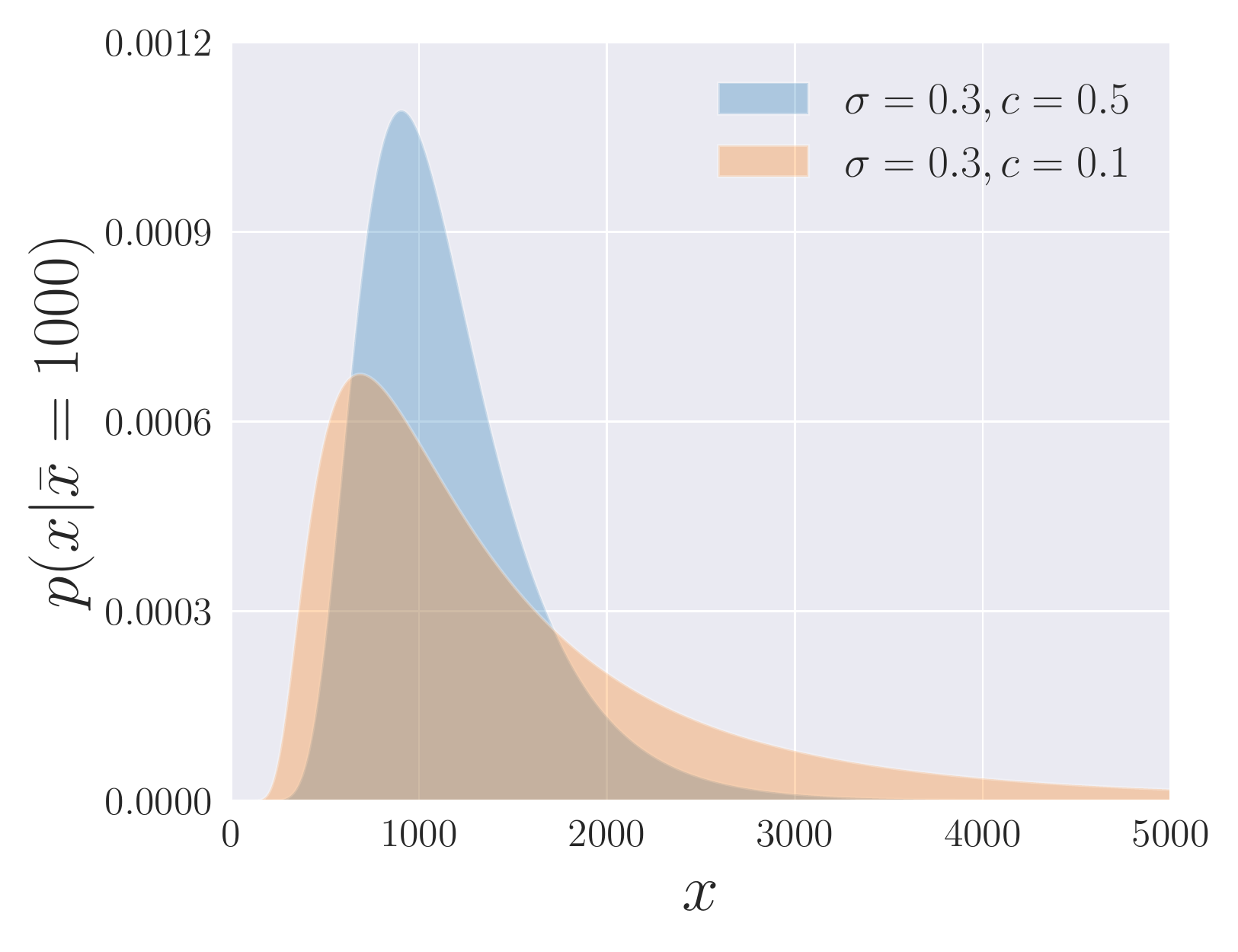}}
    \caption{Example probability distribution functions for the (a) lognormal and (b) power lognormal distributions. The expectation value for each distribution is $\bar{x} = 1000$.}
    \label{fig:pdfs}
\end{figure}

Example pdfs for the lognormal and power lognormal distributions may be found in Fig. \ref{fig:pdfs}.

\subsection{Temperature Fluctuations}
When considering the effects of temperature fluctuations, the normal and lognormal distributions are used, the latter of which is observed in the solar wind plasma \citep{2000JGR...105.2357B}. In addition, a correlation between density and temperature via a polytropic relation $(T \propto n_{e}^{\gamma - 1})$ is examined.

Since the radio continuum emission is dependent on temperature, the emission measure as defined in equation (\ref{eq:emissionMeasure}) must be modified when temperature fluctuations are present. In the radio regime ($h\nu \ll kT$) the free-free emission coefficient can be approximated as $j_\nu^\mathit{ff} \propto n_{e}^{2} / \sqrt{T}$ \citep{1986rpa..book.....R}. To retain the definitional accuracy of equation (\ref{eq:emissionMeasureDensity}), the emission measure is modified accordingly,
\begin{equation}\label{eq:emissionMeasureTemperature}
\mathrm{EM} \equiv {\sqrt{\langle T \rangle}}\int \frac{ n_{e}^{2}(s)}{ \sqrt{T(s)} } \; ds
\end{equation}
where $\langle T \rangle$ is an assumed (or estimated) temperature that can either represent the nebula as a whole or vary for each line of sight (as, for example, determined from spectroscopic line ratios).

\subsection{Emission}
The prominence of various nebular emission lines depends on the detailed balance of electronic transitions. In equilibrium, this balance of collisional and radiative transitions into and out of level $i$ is described by the expression
\begin{equation}
\sum_{j\not=i} f_j n_e q_{ji} + \sum_{j>i} f_{j} A_{ji} = \sum_{j\not=i} f_{i} n_{e} q_{ij} + \sum_{j<i} f_{i} A_{ij}
\end{equation}
where $f_{k}$ is the fraction of ions in level $k$, $q_{kl}$ is the collisional transition rate from level $k$ to level $l$, and $A_{kl}$ is the Einstein coefficient for the radiative transition from level $k$ to level $l$ \citep{2006agna.book.....O}. Given $n_{e}$ and $T$ one can solve for $f_{k}$; the emission coefficient $j_{kl}$ is then computed from
\begin{equation}
4\pi j_{kl} = f_{k} n(X^{+i}) A_{kl} h \nu_{kl}
\end{equation}
where $n(X^{+i})$ is the density of the emitting ion.

We determine the emission coefficients by generating emissivities using \textsc{pyneb} \citep{2015A&A...573A..42L}, a \textsc{python} library for analyzing emission lines of various atoms and ionization levels based on the original work by \cite{1987JRASC..81..195D} and expanded upon by \cite{1995PASP..107..896S}. The emissivity returned for a given electron density and temperature, $\varepsilon_{\lambda}(n_{e},T)$, is related to the emission coefficient, $j_{\lambda}(n_{e},T)$, by the following relation
\begin{equation}\label{eq:pynebEmiss}
j_\lambda(n_{e},T) = n_{e} n(X^{+i}) \varepsilon_\lambda(n_{e},T).
\end{equation}
With the assumption\footnote{Since nebulae are dominated by photoionisation, the Saha equation is unlikely to apply. A radial dependence of the fractional ionizational states would be expected, however, with greater ionized states occurring more frequently towards the ionizing source(s). Pockets of dense neutral concentrations have also been observed in nebulae, indicating deviations from this assumption \citep[e.g.][]{2000AJ....119.2910O, 2009ApJ...700.1067M}.} $n(X^{+i}) \propto n_{e}$, the line intensity is found by integrating the emission coefficient along the line of sight
\begin{equation}
I_{\lambda} = \int j_\lambda(s) \; ds.
\end{equation}

\subsection{Diagnostics}
Electron densities can be measured using CELs with similar excitation energies but differing collisional and/or radiative transition rates that originate from the same ion. Similar excitation rates ensures that the relative populations remain roughly constant with changing temperatures, while differing collisional / radiative rates causes the relative populations to depend on the electron density.

\begin{figure}
	\begin{center}
	\includegraphics[width=\columnwidth]{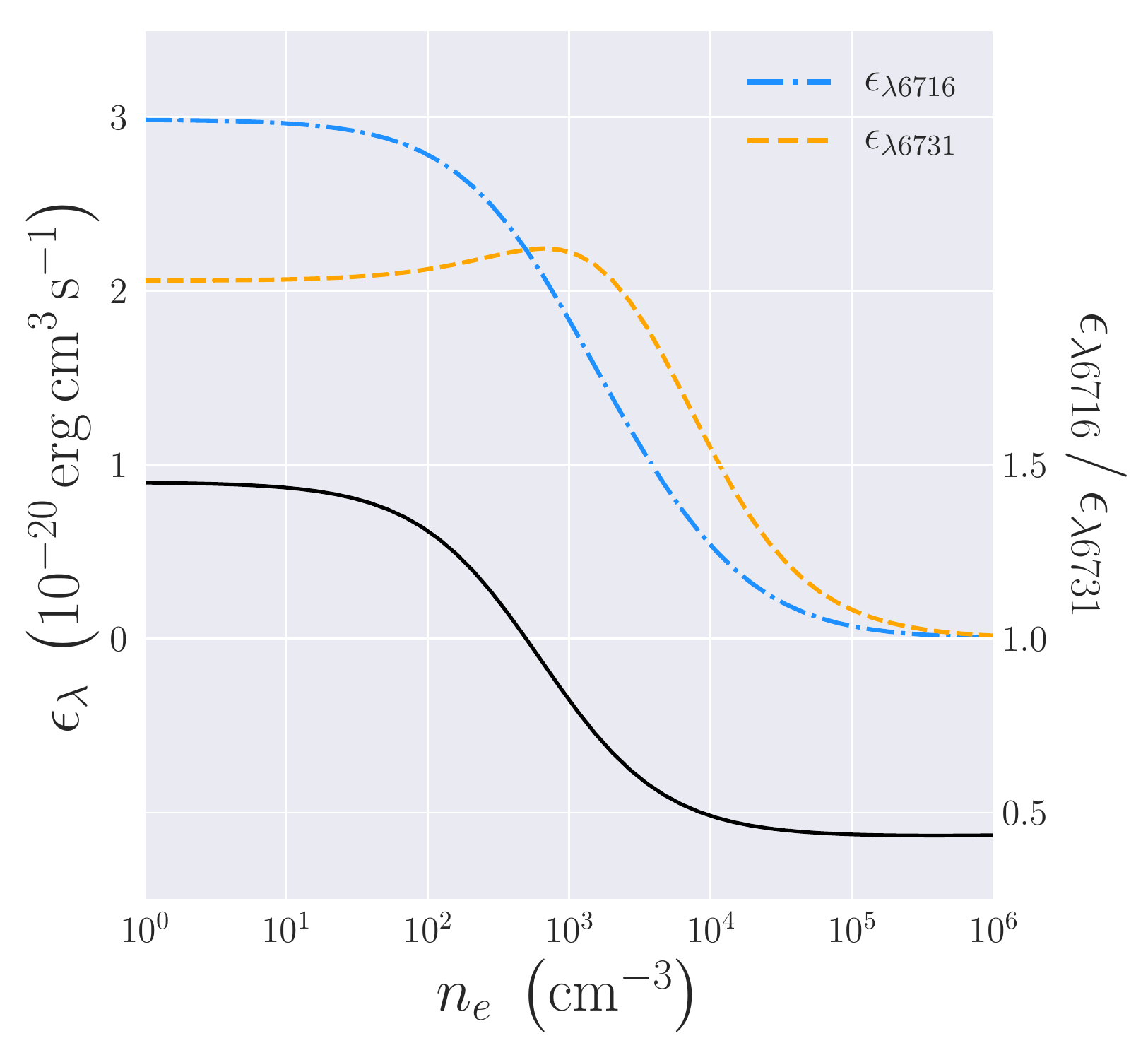}
	\caption{Emissivities of the forbidden lines [\ion{S}{ii}] $\lambda 6716, \, \lambda 6731$ as a function of the electron density at a temperature $T = 10^{4} \; \rm K$. Given the line ratio (solid black line; axis on right), one can recover the electron density.}
	\label{fig:line_ratio}
	\end{center}
	\end{figure}

One example is the [\ion{S}{ii}] $\lambda 6716$, $\lambda 6731$ set of emission lines, which are illustrated in Fig. \ref{fig:line_ratio}. The decline in emissivity with increasing electron density is caused by collisional de-excitation. The density at which de-excitation becomes important is denoted the critical density. More formally, the critical density $n_{c}$ for any level $i$ is defined as
\begin{equation}
n_{c}(i) = \frac{\sum_{j < i} A_{ij}}{\sum_{j\not=i} q_{ij}}.
\end{equation}
Similarly, the line ratio of CELs that have different excitation energies can be used to determine the temperature. Some standard line diagnostics are given in Tables \ref{tab:emission_line_density} -- \ref{tab:emission_line_temperature}.

	\begin{table}
	\begin{center}
	\caption{Emission lines used for density estimates. $n_\star$ denotes the electron density at which the line ratio is between its minimum and maximum values at a temperature $T = 10^{4} \, \mathrm{K}$ and were computed using \textsc{pyneb}.}
	\label{tab:emission_line_density}
	\begin{tabular}{lcc}
	\hline
	Ion & Line Ratio & $n_\star \; \left(\mathrm{cm}^{-3}\right)$ \\
	\hline
	{[\ion{Ar}{iv}]}  & $\lambda 4711 / \lambda 4740$ & 6620\\
	{\ion{C}{iii}]}   & $\lambda 1907 / \lambda 1909$ & 35000\\
	{[\ion{Cl}{iii}]} & $\lambda 5518 / \lambda 5538$ & 3880\\
	{[\ion{K}{v}]}    & $\lambda 4123 / \lambda 4163$ & 25700\\
	{[\ion{N}{i}]}    & $\lambda 5198 / \lambda 5200$ & 1610\\
	{[\ion{O}{ii}]}   & $\lambda 3726 / \lambda 3729$ & 710\\
  {[\ion{O}{iii}]}  & $52\mu\mathrm{m} / 88\mu\mathrm{m}$ & 210 \\
	{[\ion{S}{ii}]}   & $\lambda 6716 / \lambda 6731$ & 620\\
	\hline
	\end{tabular}
	\caption{Emission lines used for temperature estimates.}
	\label{tab:emission_line_temperature}
	\begin{tabular}{lc}
	\hline
	Ion & Line Ratio \\
	\hline
	{[\ion{N}{ii}]}   & $\lambda(6548 + 6584) / \lambda 5754$	\\
	{[\ion{Ne}{iii}]} & $\lambda (3869 + 3968) / \lambda 3343$	\\
	{[\ion{O}{iii}]}  & $\lambda(5007 + 4959) / \lambda 4363$ 	\\
	{[\ion{S}{iii}]}  & $\lambda(9531+ 9069) / \lambda 6312$	\\
	\hline
	\end{tabular}
	\end{center}
	\end{table}

Deriving the abundance requires knowledge of the electron density and temperature in order to estimate the emissivity. For ORLs, the density estimate usually comes from CELs, while the Balmer jump is often used to estimate the temperature. The abundance relative to hydrogen is then found by comparing the observed intensity to a prominent recombination line of hydrogen, such as $H{\beta}$, via the following:
\begin{equation}
\frac{X^{+i}}{H^{+}} = \frac{I(\lambda)}{I(H\beta)}\frac{\varepsilon_{H\beta}(n_e, T)}{\varepsilon_\lambda(n_e, T)}.
\end{equation}

The Balmer jump is defined as the difference between the left- and right-hand limits of the nebular continuum flux density, $I_{c}$, at the limit of the Balmer series ($\lambda_\mathcal{B}=3645$ {\AA}):
\begin{equation}
BJ \equiv I_{c}\left(\lambda_\mathcal{B}^{-}\right) -  I_{c}\left(\lambda_\mathcal{B}^{+}\right)
\end{equation}
Since the fraction of electrons in the first excited state depends on the temperature, the strength of the Balmer jump serves as a temperature diagnostic, with the strength going as $BJ \propto n_{e}^{2} \, T^{-3/2}$ \citep{2011piim.book.....D}.

To derive the temperature from the Balmer jump, a comparison is usually done with a Balmer recombination line, typically the $n = 11\rightarrow 2$ transition at 3770 {\AA} \citep{2001MNRAS.327..141L}. We therefore treat the Balmer jump as an emission line (that is, $j_{BJ} = n_{e}^{2} T^{-3/2}$) and generate the H11 emission using \textsc{pyneb}. Given the electron density and the observed line ratio, we can recover the temperature, which is done using an interpolation spline.

\subsection{Grid Size}

The characteristic cell size is an important simulation parameter, the choice of which affects the variance of the observed parameters. Increasing the number of cells lessens the impact of any individual cell, causing the observed properties of each line of sight to approach the statistical mean of the distribution, in accordance with the central limit theorem.\footnote{A Pareto-like distribution of densities can be an exception, with more cells leading to a greater density estimate. The number of cells, however, mainly affects the \emph{spread} of the observed filling factors and ADFs.}

\begin{figure}
\begin{center}
\includegraphics[width=\columnwidth]{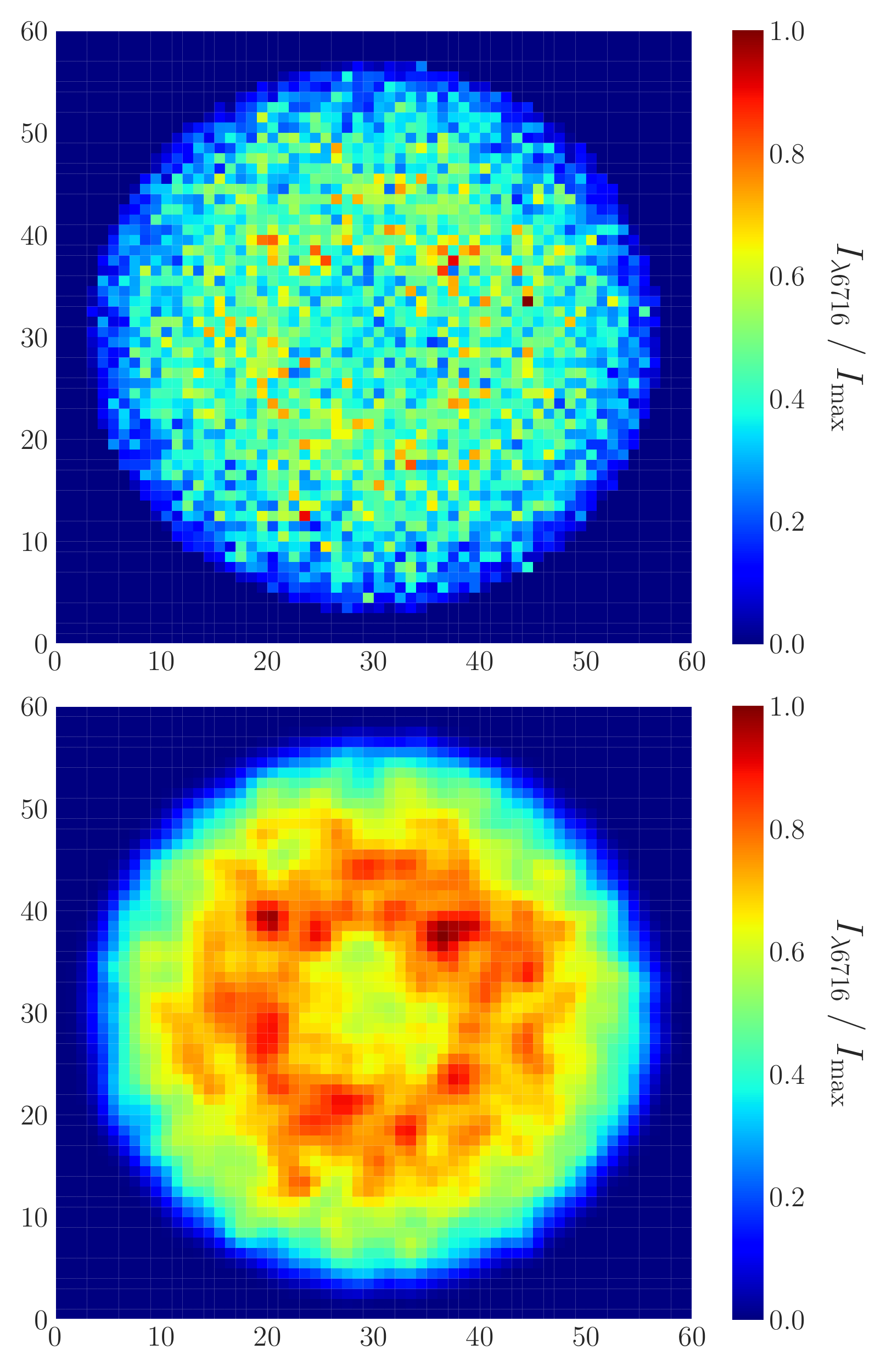}
\caption{Normalized [\ion{S}{ii}] $\lambda 6716$ intensity for a constant temperature nebula $(T=10^{4} \, \mathrm{K})$, where the density in each cell was drawn from an exponential distribution with mean $\bar{n}_{e} = 1000 \, \mathrm{cm}^{-3}$. The inner radius is $\frac{1}{3}$ of the outer radius. The bottom image was produced by convolving the top image using a Gaussian filter with a one standard deviation kernel.}
\label{fig:convolution}
\end{center}
\end{figure}

When the number of cells is large $(\gtrsim 100^{3})$, simulated nebulae take on a fairly uniform, yet grainy, appearance where the brightness closely mirrors the depth. Inhomogeneities are observed when using less cells, but the intensity images become pixelated. This can be remedied by convolving with a Gaussian filter, the properties of which are similar to increasing the cube size at a reduced computational cost, but also resulting in a more `realistic' image (Fig. \ref{fig:convolution}). The intensity images are convolved with a one standard deviation kernel unless otherwise noted.

Convolving, however, can greatly affect the EM density estimates for those lines of sight only a few cells deep, which consequently show up as outliers. Proceeding with the geometry in Fig. \ref{fig:geomSphere} ($30^{3}$ cells), these outliers can be removed by requiring at least 6 cells along the line of sight, which leaves 548 lines of sight through each simulated nebula.

While it might be argued that the resolution is rather coarse, in turbulence the largest amplitude fluctuations (of magnetic field, density, etc.) occur at the outer scale, which is the largest turbulent scale. Our calculations would imply that the outer scale is $\sim 10\%$ of the diameter of the region. This choice has some basis from studies of turbulence in other media, such as fluid flow and the solar wind. The largest eddies in turbulent channel flow have sizes comparable to the boundary layer or shear layer thickness, which is roughly a fraction of the transverse dimension of the channel \citep{rouse1946elementary, 1978turb.book.....B}. Fluid turbulence in a channel or pipe flow has the advantage that the scale size of the system is obvious. This scale is less obvious in an open astrophysical medium without rigid boundaries, however.

An excellent model for astrophysical turbulence is the well-studied turbulence that exists in the solar wind. Analyzing remote sensing observations of the solar wind at heliocentric distances of $15 - 25 \, R_\odot$, \cite{2002ApJ...576..997S} adopted an outer scale of $\sim 1 \, R_{\odot}$ to infer the magnitude of density fluctuations. \cite{2001SSRv...97....9W} also used radio propagation studies, although of a different type, to infer properties of the turbulence in the same range of heliocentric distance. They measured a feature in the power spectrum of radio beacon frequency fluctuations which could be interpreted as the outer scale of turbulence, and which yielded a value of $\sim 3 \, R_{\odot}$ at a heliocentric distance of $20 \, R_{\odot}$. Adopting these distances as the `system size', this would imply an outer scale that is $\sim 10\%$ of the system diameter.

Another astrophysical plasma with well-known turbulent properties is the interstellar medium (ISM), most specifically the Warm Ionized Medium (WIM). \cite{2013SSRv..178..483H} provide a review of turbulence in the ISM, noting that some measurements of the WIM indicate an outer scale of about 5 pc, while others give larger values of 50 - 100 pc. Both values may be correct, with the former referring to the fully three-dimension turbulence, while the latter may be more accurate for planar, two-dimensional eddies. The most obvious value to use for the system size is the thickness of the layer containing the WIM gas, which is 2 - 3 kpc \citep{2009RvMP...81..969H}. An outer scale of 5 pc would then be considerably finer than our simulations, while an outer scale of 50 - 100 pc would be $\le 5\%$ of the system size.
\section{Simulation Results}\label{s:simulations}

\subsection{Density Fluctuations}\label{s:den}
The most straightforward explanation for the non-unity value of the filling factor would be the presence of density fluctuations. Indeed, such fluctuations naturally lead to discrepancies between the estimated densities from the emission measure and CELs. The magnitude of the discrepancy, however, depends on the underlying distribution of electrons, both in the pdf and the mean density relative to the critical densities.

\begin{figure}
\centering
\includegraphics[width=\columnwidth]{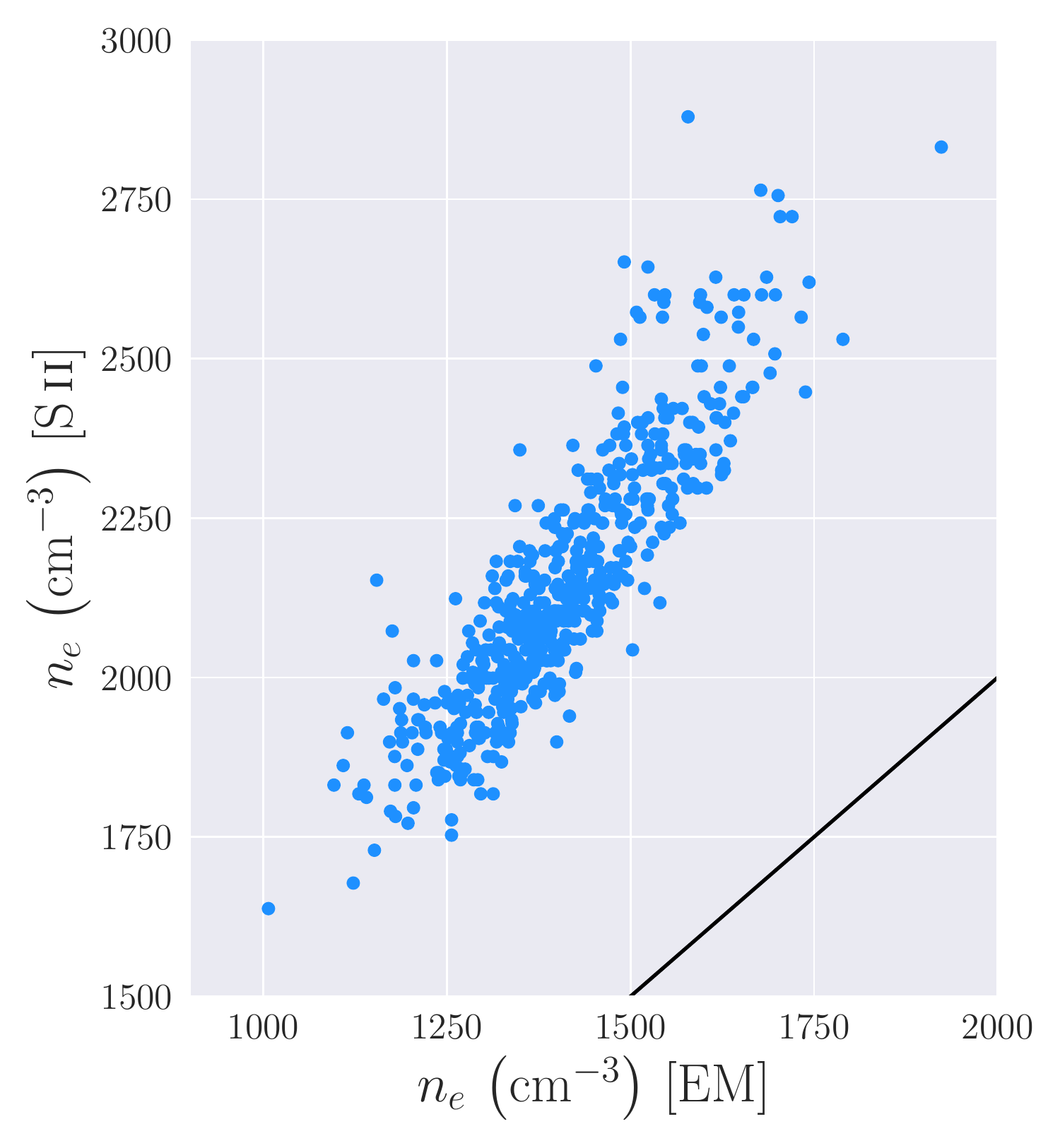}
\caption{Derived densities for 548 lines of sight using the observed \textsc{[S ii]} $\lambda 6716 / \lambda 6731$ line ratio and the emission measure. The temperature was set to $T = 10^{4} \; \mathrm{K}$, while the densities were drawn from an exponential distribution with mean $\bar{n}_{e} = 1000 \; \mathrm{cm}^{-3}$. The solid black line represents the 1:1 ratio.}
\label{fig:logScatter}
\captionof{table}{Average line-of-sight density estimates for an exponentially density-distributed nebula with mean $\bar{n}_{e} \approx 1000 \, \rm cm^{-3}$; $\sigma$ represents the standard deviation of the observed values.}
\begin{tabular}{lcc}
\hline
Diagnostic & $\langle n_{e} \rangle \; \rm \left(cm^{-3}\right)$ & $\sigma \; \rm \left(cm^{-3}\right)$ \\
\hline
Emission Measure                                & 1410  & 130 \\
{[\ion{Ar}{iv}]}  $\lambda 4711 / \lambda 4740$ & 2870  & 410 \\
{\ion{C}{iii}]}   $\lambda 1907 / \lambda 1909$ & 3010  & 480 \\
{[\ion{Cl}{iii}]} $\lambda 5518 / \lambda 5538$ & 2720  & 350 \\
{[\ion{K}{v}]}    $\lambda 4123 / \lambda 4163$	& 2990  & 470	\\
{[\ion{N}{i}]}    $\lambda 5198 / \lambda 5200$ & 2020  & 180	\\
{[\ion{O}{ii}]}   $\lambda 3726 / \lambda 3729$ & 2220  & 220 \\
{[\ion{O}{iii}]}  $52\mu\mathrm{m} / 88\mu\mathrm{m}$ & 1760 & 140 \\
{[\ion{S}{ii}]}   $\lambda 6716 / \lambda 6731$ & 2140  & 200 \\
\hline
\end{tabular}
\label{tab:expden}
\end{figure}

These discrepancies are readily evident in Table \ref{tab:expden} and Fig. \ref{fig:logScatter} for the case of an exponential pdf with a constant temperature $T = 10^{4} \, \mathrm{K}$. While the true mean density is $\bar{n}_{e} \approx 1000 \, \mathrm{cm}^{-3}$, using either [\ion{S}{ii}] or [\ion{O}{ii}] as a diagnostic gives an estimate roughly twice this value, with [\ion{Cl}{iii}] and [\ion{Ar}{iv}] providing even larger estimates due to their greater critical densities. The emission measure gave the closest estimate to the true value in this instance, with $\langle {n}_{e,\mathrm{EM}} \rangle \approx 1410 \pm 130 \, \mathrm{cm}^{-3}$.

\begin{figure*}
\centering
\subfloat[Lognormal]{\includegraphics[width=\columnwidth]{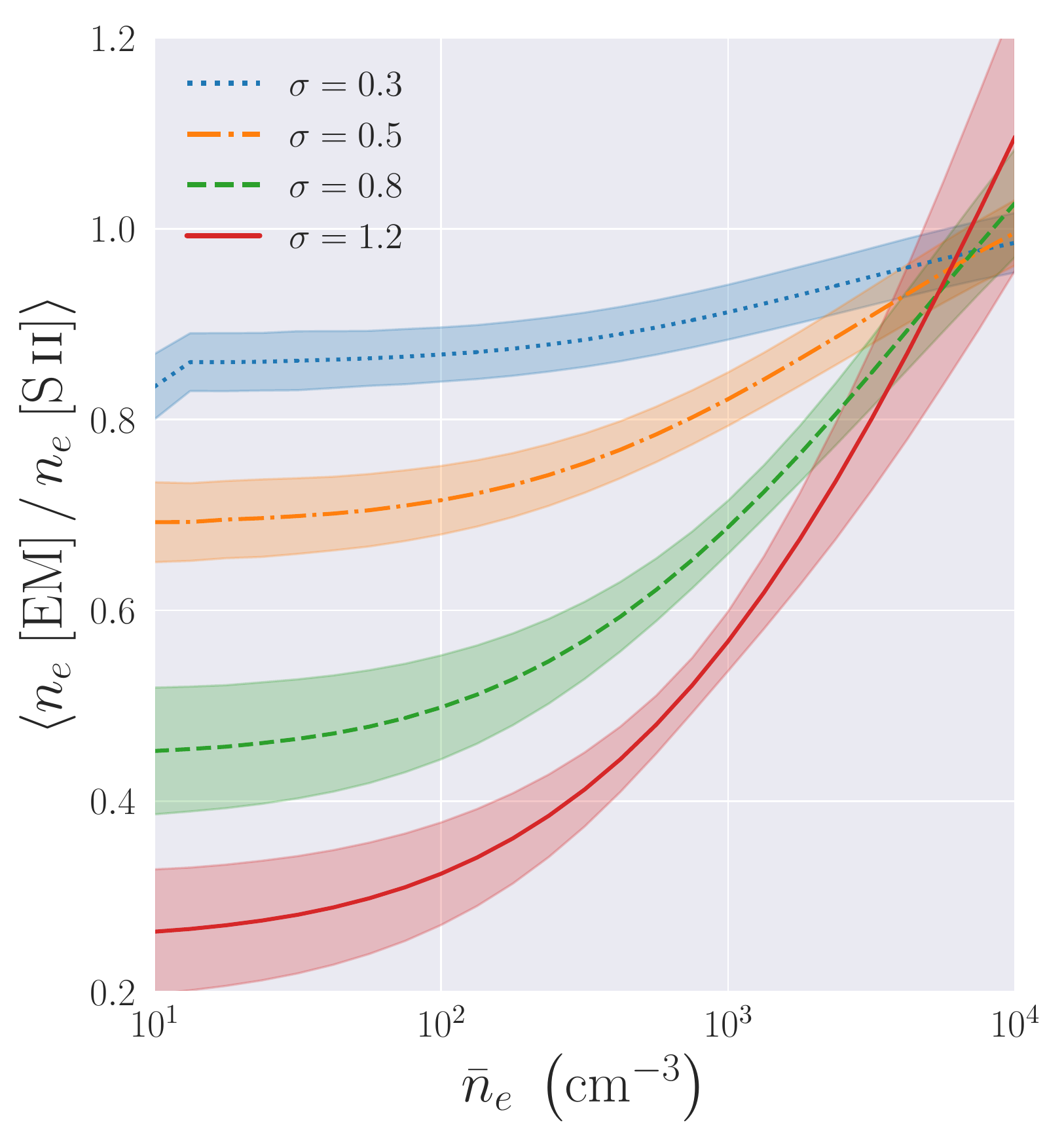} \label{fig:ff_vs_den_lognormal}}
\subfloat[Power Lognormal]{\includegraphics[width=\columnwidth]{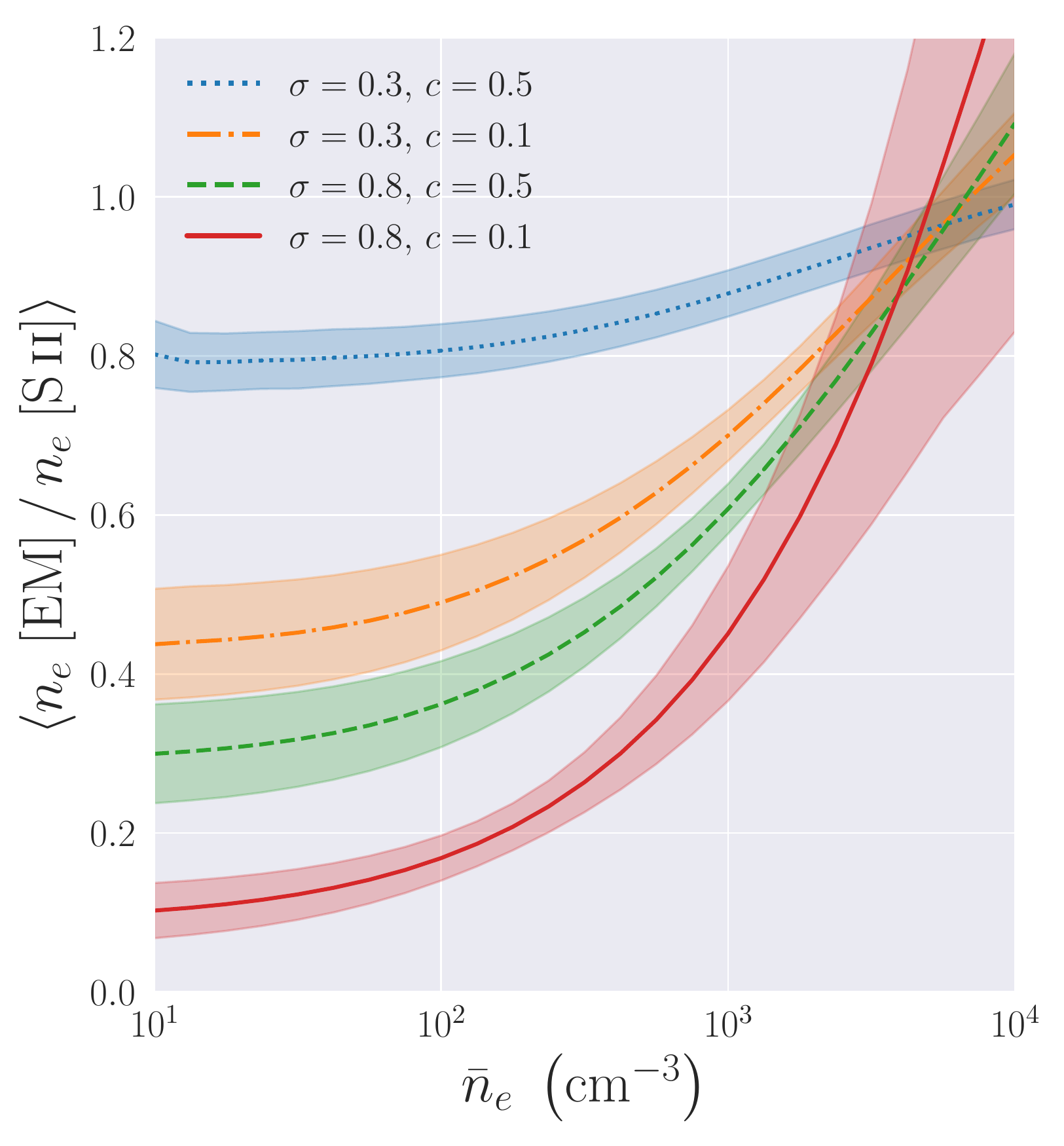}\label{fig:ff_vs_den_pareto}}\\
\caption{Average line of sight ratio of density estimates as a function of the mean nebular density $\bar{n}_{e}$ for the (a) lognormal and (b) power lognormal distributions as determined from the [\ion{S}{ii}] $\lambda 6716 / \lambda 6731$ line ratio. A constant temperature $T=10^{4} \, \mathrm{K}$ was used. The ribbon represents the one standard deviation spread of the observed values.}
\label{fig:ff_vs_den}
\end{figure*}

Our simulation results for a lognormal and power lognormal distribution of densities can be found in Figs. \ref{fig:ff_vs_den_lognormal} and \ref{fig:ff_vs_den_pareto} respectively, where the ratio of density estimates is plotted as a function of the mean nebular density. Greater discrepancies (corresponding to smaller filling factors) are observed at smaller densities and for pdfs with more extended tails, while the effect of using different ions mainly causes a shift to the left or right depending on the critical densities.\footnote{An exception for the ions listed in Table \ref{tab:expden} are the semi-forbidden {\ion{C}{iii}]} $\lambda 1907, \lambda 1909$ lines. The profile displays a typical sigmoidal (or $\mathcal{S}$-shaped) curve, with the ratio asymptotically approaching limits below unity in the low- and high-density regimes. The profiles of the forbidden lines suggest a possible sigmoidal curve as well, before the onset of numerical issues.}

The origin of this discrepancy comes from the emissivity dependence of the emission lines on the electron density caused by collisional de-excitation. Take, for instance, the [\ion{S}{ii}] lines, which are illustrated in Fig. \ref{fig:line_ratio}. Initially, in the case of low electron densities, the emissivity $\varepsilon_{\lambda 6716}$ declines with increasing density while $\varepsilon_{\lambda 6731}$ stays relatively constant. Thus the densest and brightest regions along the line of sight will emit disproportionately less light at $\lambda 6716$ compared to what the mean density would suggest, decreasing the line ratio $I_{\lambda 6716} / I_{\lambda 6731}$ and inflating the density estimate. Once $\varepsilon_{\lambda 6731}$ begins to decline these tend to offset, bringing the density estimate more in line with the emission measure.

The filling factors reported here are corroborated by an approximate analytic theory for the case of an exponential pdf, while confirming the result reported here that smaller filling factors are inferred for lower mean densities relative to the critical density \citep{2019arXiv191008466S}.

A general result that emerges is that in the case of density fluctuations, a higher estimate for the electron density will be obtained from line doublets with larger critical densities (roughly represented by $n_\star$ in Table \ref{tab:emission_line_density}), although the estimates converge in the low-density regime. This prediction is at least qualitatively consistent with some of the published results concerning PNe. \cite{1989ApJ...343..811S} found that densities estimated from [\ion{O}{ii}] and [\ion{S}{ii}] were generally lower than those obtained from [\ion{Cl}{iii}] and [\ion{Ar}{iv}]. \cite{2002A&A...382..282C} similarly found that [\ion{O}{ii}] gave lower estimates than [\ion{Cl}{iii}] and [\ion{Ar}{iv}], but found the [\ion{S}{ii}] estimates to be more comparable to the latter. \cite{2004A&A...427..873W} found that densities estimated from the [\ion{Ar}{iv}] lines were higher than the estimates from [\ion{S}{ii}] and [\ion{O}{ii}], particularly for nebulae with higher densities.  However, the densities estimated from [\ion{Cl}{iii}] were generally consistent with those obtained from [\ion{S}{ii}] and [\ion{O}{ii}]. A quantitative comparison of nebular models possessing density fluctuations with the aforementioned observations would be a worthwhile and interesting investigation.

\begin{figure*}
\subfloat[Lognormal]{\includegraphics[width=\columnwidth]{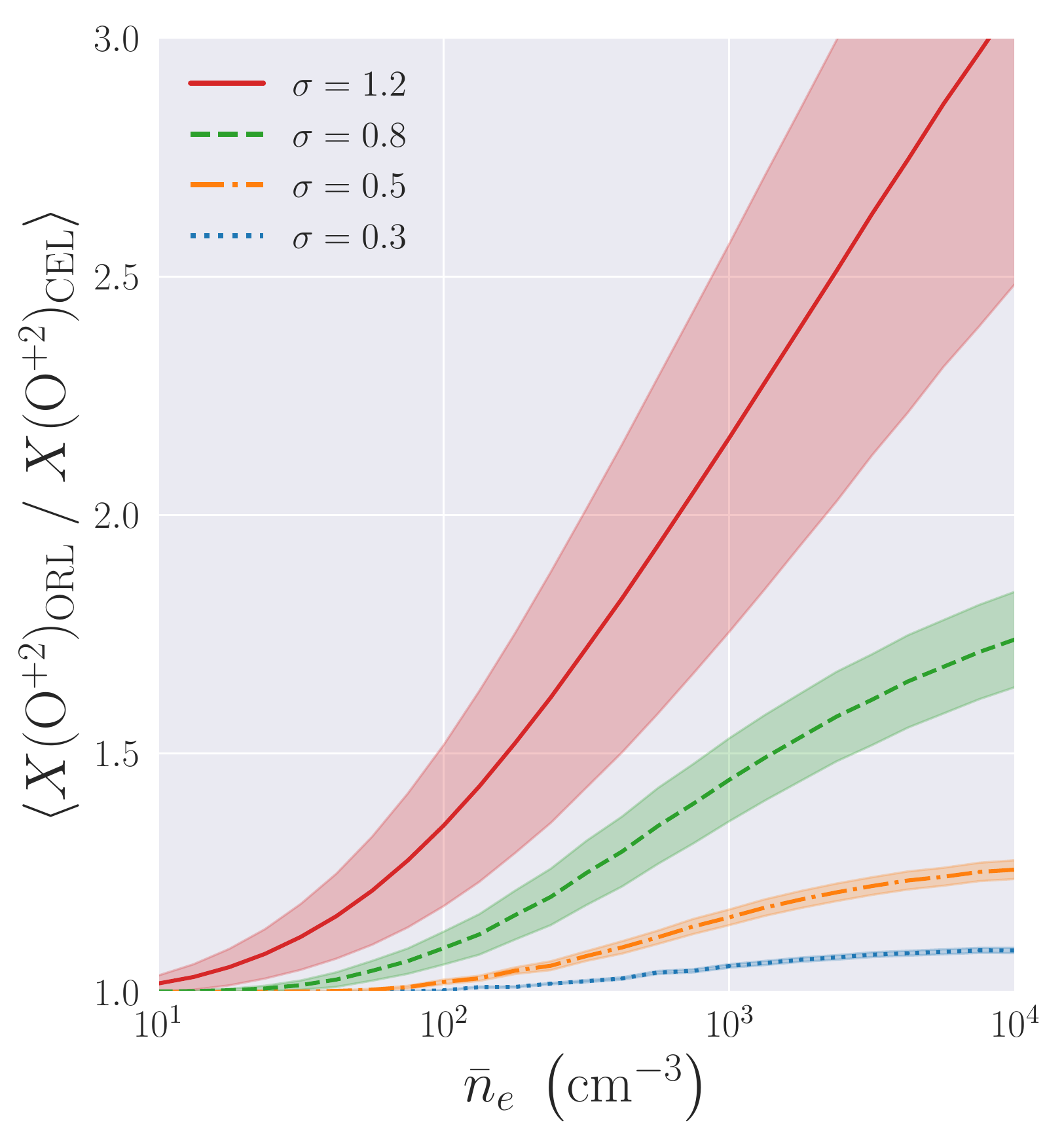}}
\subfloat[Power Lognormal]{\includegraphics[width=\columnwidth]{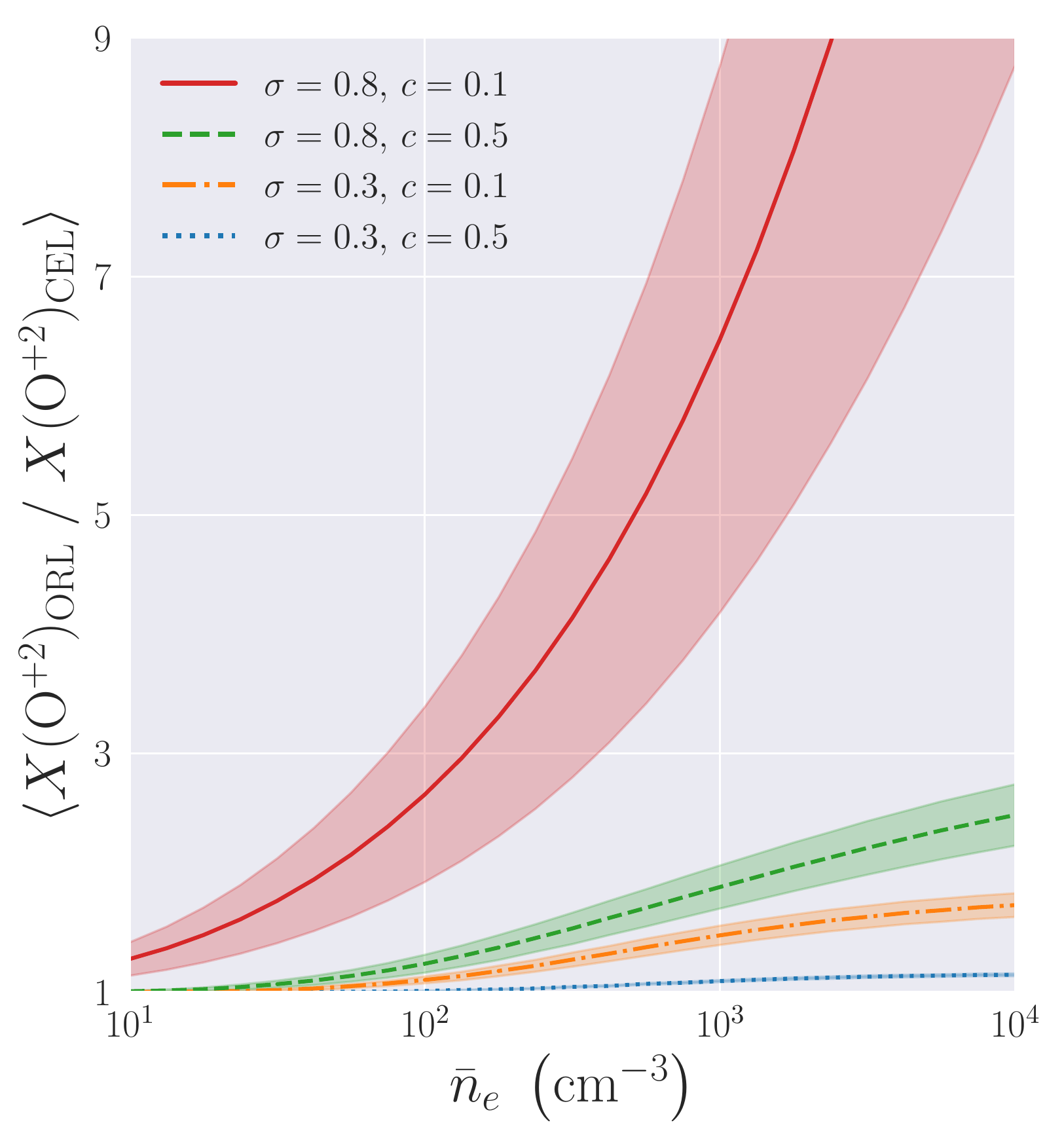}}
\caption{Average line of sight ADF as a function of the mean nebular density $\bar{n}_{e}$ for the (a) lognormal and (b) power lognormal distributions. A constant temperature $T=10^{4} \, \mathrm{K}$ was used. The electron density and CEL abundance were derived from the [\ion{O}{iii}] lines at 52 and 88 $\mu\rm m$; the ORL abundance from the \ion{O}{ii} recombination line at 4089 {\AA}. The ribbon represents the one standard deviation spread of the observed values. Since the distribution of ADFs are highly skewed towards larger values when significant density fluctuations are present (corresponding to the brightest regions), the convolution kernel was increased to two standard deviations.}
\label{fig:adf_vs_den}
\end{figure*}

Density fluctuations can also lead to ADFs exceeding unity when using far-infrared CELs whose emissivities are density-dependent, as shown in Fig. \ref{fig:adf_vs_den}. An increase in density leads to a decrease in the emissivity of CELs, while the emissivities of ORLs stay relatively constant. Thus ORLs will appear disproportionately bright compared to CELs, resulting in a greater estimate for the abundance. Since the emissivity of ORLs better traces the emissivity of the hydrogen recombination lines by which the hydrogen abundance is determined, the abundance estimates from ORLs are generally more accurate than those from CELs, whose accuracy generally declines with increasing electron density and is the prominent reason for the density-dependence of the ADF. This becomes particularly prominent for a Pareto-like distribution of electron densities, where the distribution of ADFs can be highly skewed towards larger values. We therefore increased the convolution kernel to two standard deviations to create a more symmetric distribution of ADFs.

While it appears from Fig. \ref{fig:adf_vs_den} that a Pareto-like distribution of electron densities is capable of generating large ADFs, the derived electron densities are significantly larger than the mean density, pushing the spectroscopic line ratio to the limit of its applicability. For example, even though the mean density is $\bar{n}_{e} = 100 \, \mathrm{cm}^{-3}$, the mean density along each line-of-sight inferred from [\ion{O}{iii}] under a power lognormal distribution with parameters $\sigma = 0.8$, $c = 0.1$ is around $\langle n_{e} \rangle \approx 1800 \, \mathrm{cm}^{-3}$, with a comparable spread in values $\sigma \approx 1100 \, \mathrm{cm}^{-3}$. Thus abundance discrepancies only up to factors of a few ($ADF \lesssim 5$) might be accounted for by density fluctuations when using far-infrared CELs.

When using optical CELs whose emissivities are nearly independent of the electron density (e.g. [\ion{O}{iii}] $\lambda 5007$) the observed ADFs are close to one. Thus density fluctuations are not a viable explanation for resolving the abundance discrepancy problem, but could be a contributing factor to the abundance discrepancies when using far-infrared CELs.

For PNe, it is generally the case that far-infrared and optical CELs give similar abundances, even for the more moderate ADFs \citep{2004MNRAS.353.1251L, 2004MNRAS.353..953T}. From Fig. \ref{fig:adf_vs_den} we would expect larger ADFs to be associated with higher densities, which runs contrary to the observational results presented by \cite{2018MNRAS.480.4589W}; see their fig. 16. For PNe with binary central stars, there is an inverse empirical relationship between the electron density and the ADF. This conclusion may extend to PNe in general, at least among those with more extreme ADFs, although the authors favor the view that `normal' PNe have low ADFs over a wide range of densities while the inverse relationship is restricted to binary systems. For \ion{H}{ii} regions, there appears to be no obvious dependence of optical ADFs on the density, but given the scatter in the measurements, a weak positive dependence on density cannot be excluded, which would be expected if far-infrared CELs give similar abundances. We do note, however, that in the presence of density fluctuations the density inferred from CELs would be biased towards the denser regions along the line of sight and require knowledge of the underlying pdf in order to recover the mean density.

While density fluctuations appear to be a promising explanation for the occurrence of a below-unity filling factor, they are less favorable at explaining the observed ADFs (particularly when using optical CELs whose emissivities are relatively independent of density), while also failing to explain the differing temperatures derived from CELs and the Balmer jump. Therefore, other explanations must be taken into account.

\begin{figure*}
\subfloat[Normal, $\sigma = 0.2 \cdot\overline{T}$]{\includegraphics[width=\columnwidth]{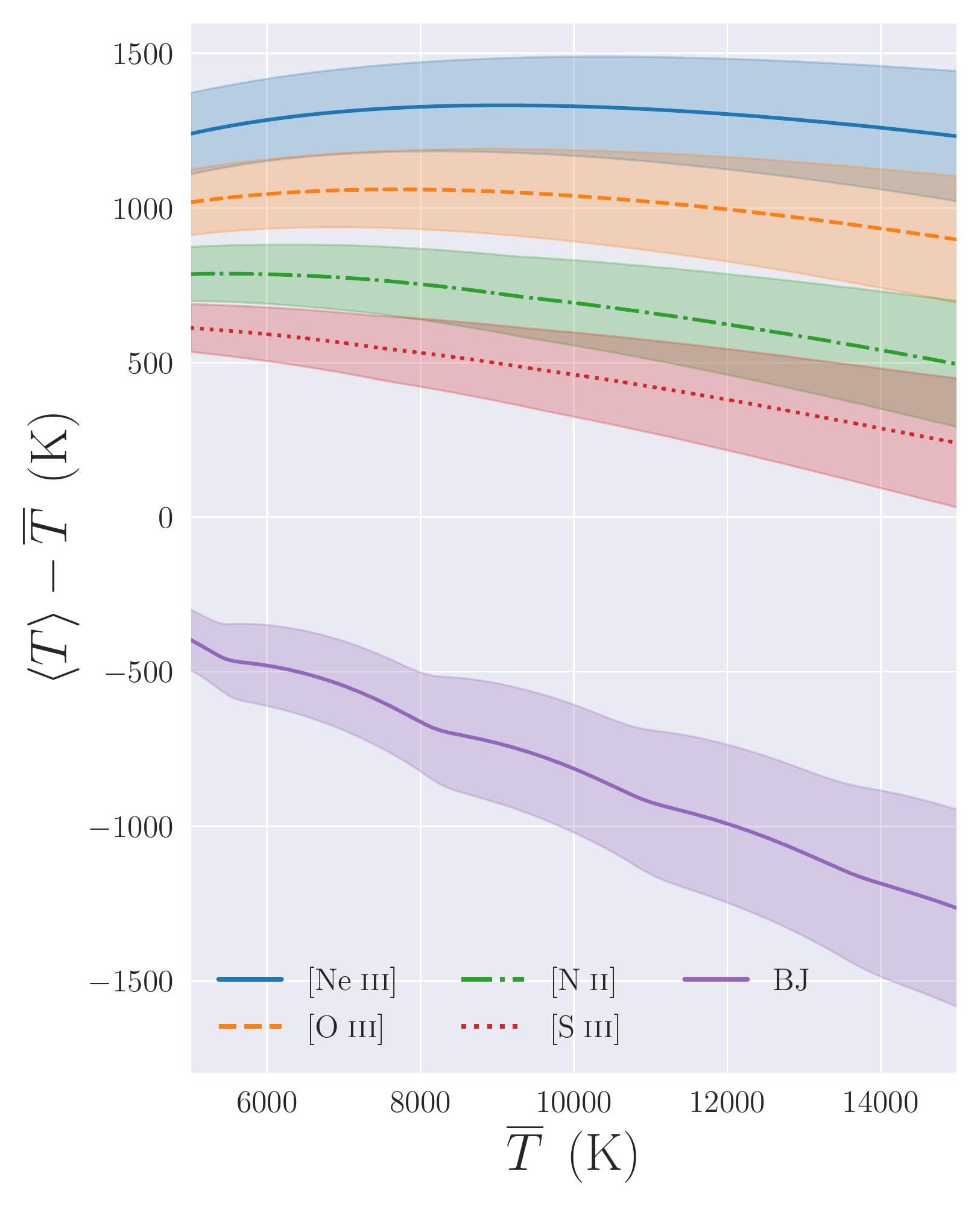}}
\subfloat[Lognormal, $\sigma = 0.3$]{\includegraphics[width=\columnwidth]{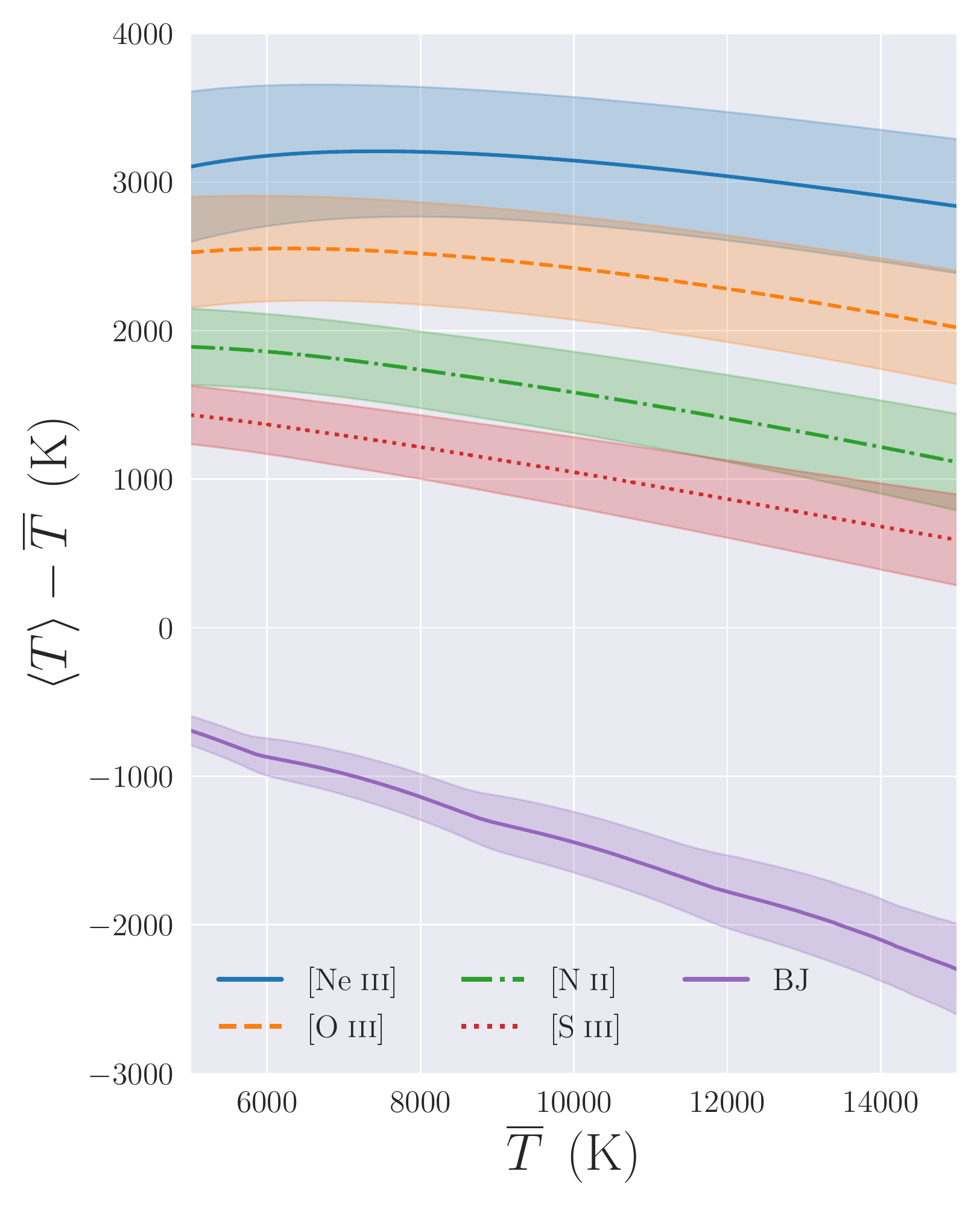}}
\caption{Average line of sight temperature estimate $\langle T \rangle$ as a function of the mean nebular temperature $\overline{T}$ for a constant density nebula $(n_{e} = 1000 \; \mathrm{cm}^{-3})$ with (a) normally- and (b) lognormally-distributed temperatures. For the normally-distributed case, the standard deviation is 20\% of the mean temperature $\overline{T}$. The ribbon represents the one standard deviation spread of the observed values. The oscillation pattern for the Balmer jump arises from \textsc{pyneb}'s use of interpolation to calculate the emissivities of ORLs.}
\label{fig:tem_tem}
\end{figure*}

\begin{figure*}
\subfloat[{[\ion{O}{iii}] 5007 {\AA}}]{\includegraphics[width=\columnwidth]{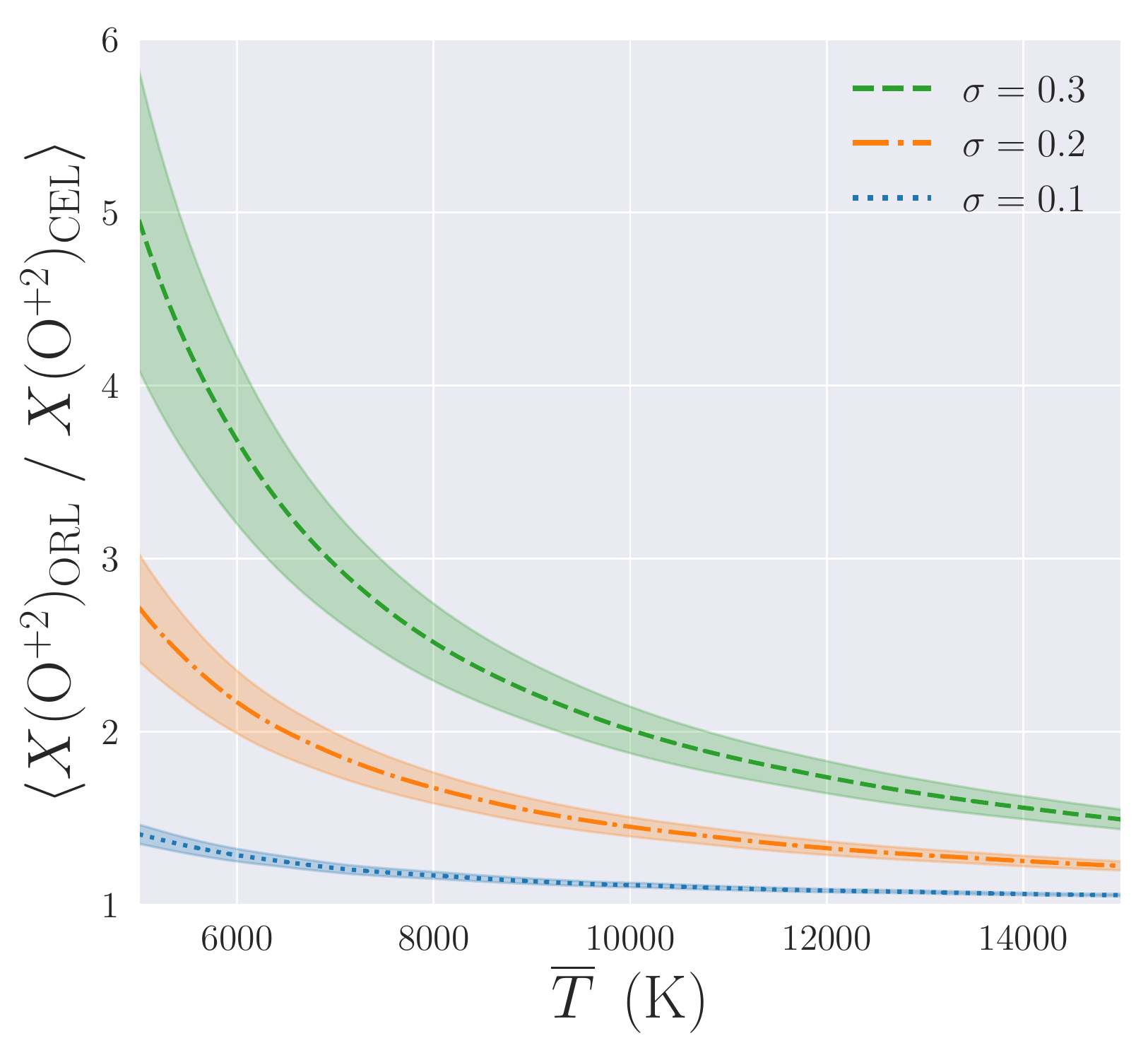}\label{fig:tem_adf_a}}
\subfloat[{[\ion{O}{iii}] 88 $\mu$m}]{\includegraphics[width=\columnwidth]{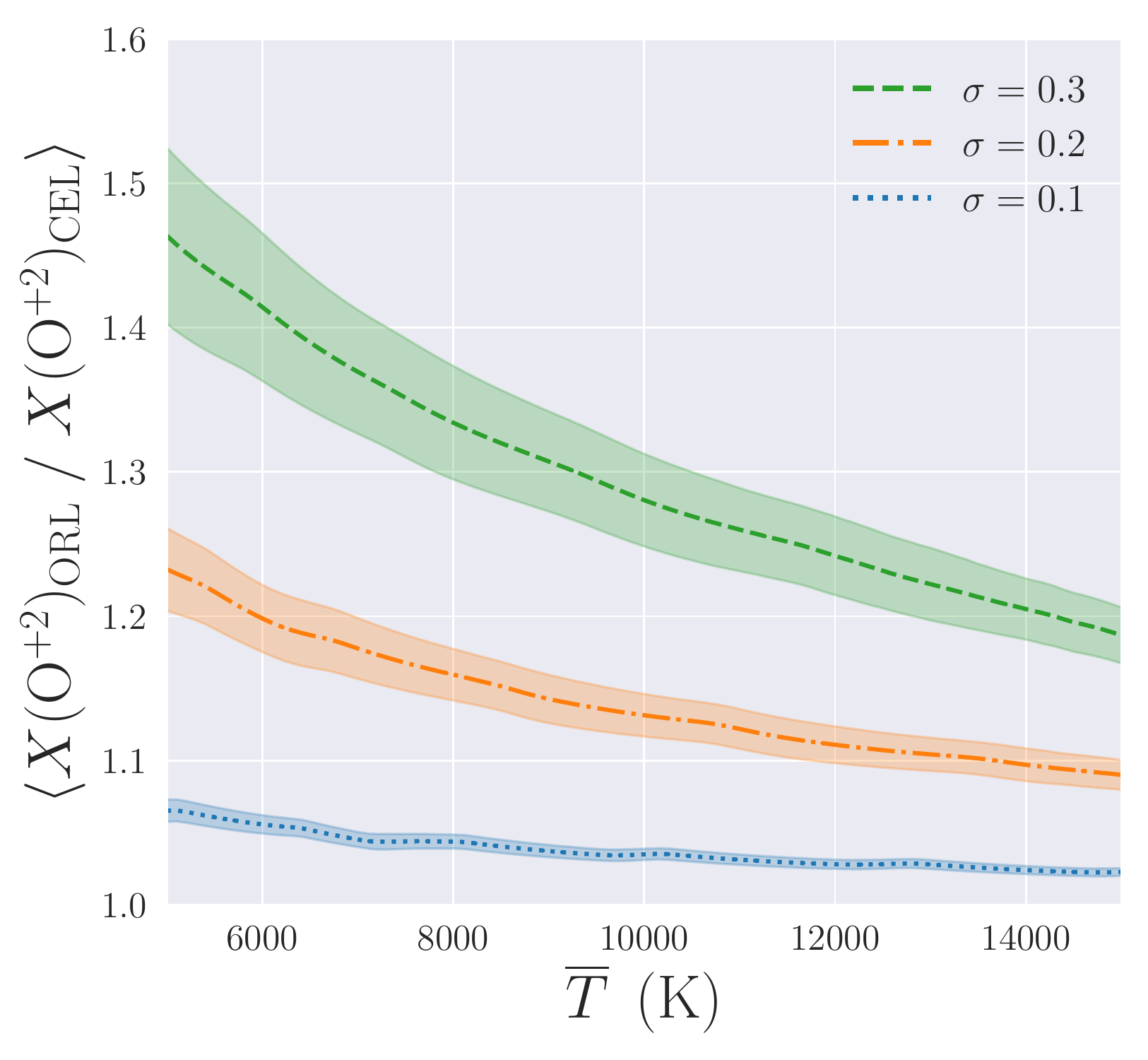}}
\caption{Average line of sight ADF as a function of the mean temperature $\overline{T}$ for a lognormal distribution of temperatures. A constant density $n_{e}=1000 \, \mathrm{cm}^{-3}$ was used. The abundances were derived from the \ion{O}{ii} $\lambda 4089$ ORL, and the (a) [\ion{O}{iii}] $\lambda 5007$ and (b) [\ion{O}{iii}] 88 $\mu$m CELs. The ribbon represents the one standard deviation spread of the observed values.}
\label{fig:tem_adf}
\end{figure*}

\subsection{Temperature Fluctuations}
Temperature fluctuations would be a straightforward explanation for the discrepancies between the Balmer jump and CELs, with the Balmer jump weighted more towards the cooler regions along the line of sight and CELs towards the warmer \citep{1967ApJ...150..825P}. They have also been put forth as the root cause for the existence of ADFs exceeding unity, particularly when estimating abundances using optical CELs. Their existence, however, is currently difficult to understand within the context of photoionisation, although radial temperature gradients are to be expected. Keeping Holmes' adage in mind,\footnote{\emph{When you have eliminated all which is impossible, then whatever remains, however improbable, must be the truth.}} we proceed to examine the effects of temperature-only fluctuations before considering the more general case of density and temperature fluctuations in \S\ref{s:den_tem}.

Fig. \ref{fig:tem_tem} shows the temperature estimates from several CELs and the Balmer jump under a normal and lognormal distribution of temperatures. Temperature fluctuations lead to biases in the estimated temperature, with greater fluctuations leading to more severe discrepancies. For CELs the magnitude of the discrepancy tends to decrease with increasing temperature as the emissivity increases at a decreasing rate. For the Balmer jump, however, the magnitude of the discrepancy tends to increase with the mean temperature, although the error relative to the mean temperature remains roughly constant in our approximation.

Perhaps the most interesting feature is the bias of the various CELs. The difference in temperature derived from [\ion{S}{iii}] and [\ion{Ne}{iii}] is comparable to the difference between [\ion{S}{iii}] and the Balmer jump; the temperature estimates from [\ion{O}{iii}] exceed those of [\ion{S}{iii}] and [\ion{N}{ii}] (figs. 5.2 and 5.3 of \cite{2006agna.book.....O} provide some observational support for this). This result suggests that a comparison of temperatures inferred from several temperature diagnostics could also provide a diagnostic for temperature fluctuations.

Significant temperature fluctuations are capable of generating large ADFs, as illustrated in Fig. \ref{fig:tem_adf}. The magnitude, however, tends to decrease with increasing mean temperature, and only moderately affects far-infrared CELs. While the Orion Nebula is observed to display a similar pattern with far-infrared CELs producing a greater abundance estimate than optical CELs \citep{2016arXiv161203633E}, this result runs contrary to observations of PNe.

ADFs are known to be largely uncorrelated with the temperature derived using CELs, but are correlated with the discrepancy between temperatures derived from the Balmer jump/ORLs and CELs \citep{2001MNRAS.327..141L, 2004MNRAS.353.1251L, 2007ApJ...670..457G}. While temperature fluctuations would predict a temperature dependence of the ADF, what is measured is not the mean temperature as CELs will be biased towards the hotter regions along the line of sight, inflating the temperature estimate and blurring the expected signature. The temperature discrepancy between the Balmer jump and CELs, meanwhile, would be an indicator of the magnitude of the temperature fluctuations, and thus a greater discrepancy would be expected to give rise to larger ADFs.

Similar results are observed when there is a temperature gradient in the absence of a central cavity. The prominent difference is that the temperature estimates, the temperature discrepancies between CELs and the Balmer jump, and the ADF all decline with increasing distance from the centre of the nebula. Since the estimated emission is a weighted average across the line of sight, the central temperatures can be significantly underestimated.

The filling factor, despite the temperature dependence of the emission measure, is relatively unaffected by the presence of temperature-only fluctuations, producing roughly equal density estimates.

While temperature fluctuations are a promising explanation for the discrepancies between CELs and the Balmer Jump, as well as the more normally distributed ADFs when using optical CELs, they are less favorable at explaining ADFs obtained from far-infrared CELs, whose emissivities are relatively independent of temperature in the range of nebular conditions. Furthermore, they are unable to explain the existence of small filling factors. Thus nebulae likely have a combination of density and temperature fluctuations -- and possibly metallicity.

\subsection{Density and Temperature Fluctuations}\label{s:den_tem}

\begin{figure}
\includegraphics[width=\columnwidth]{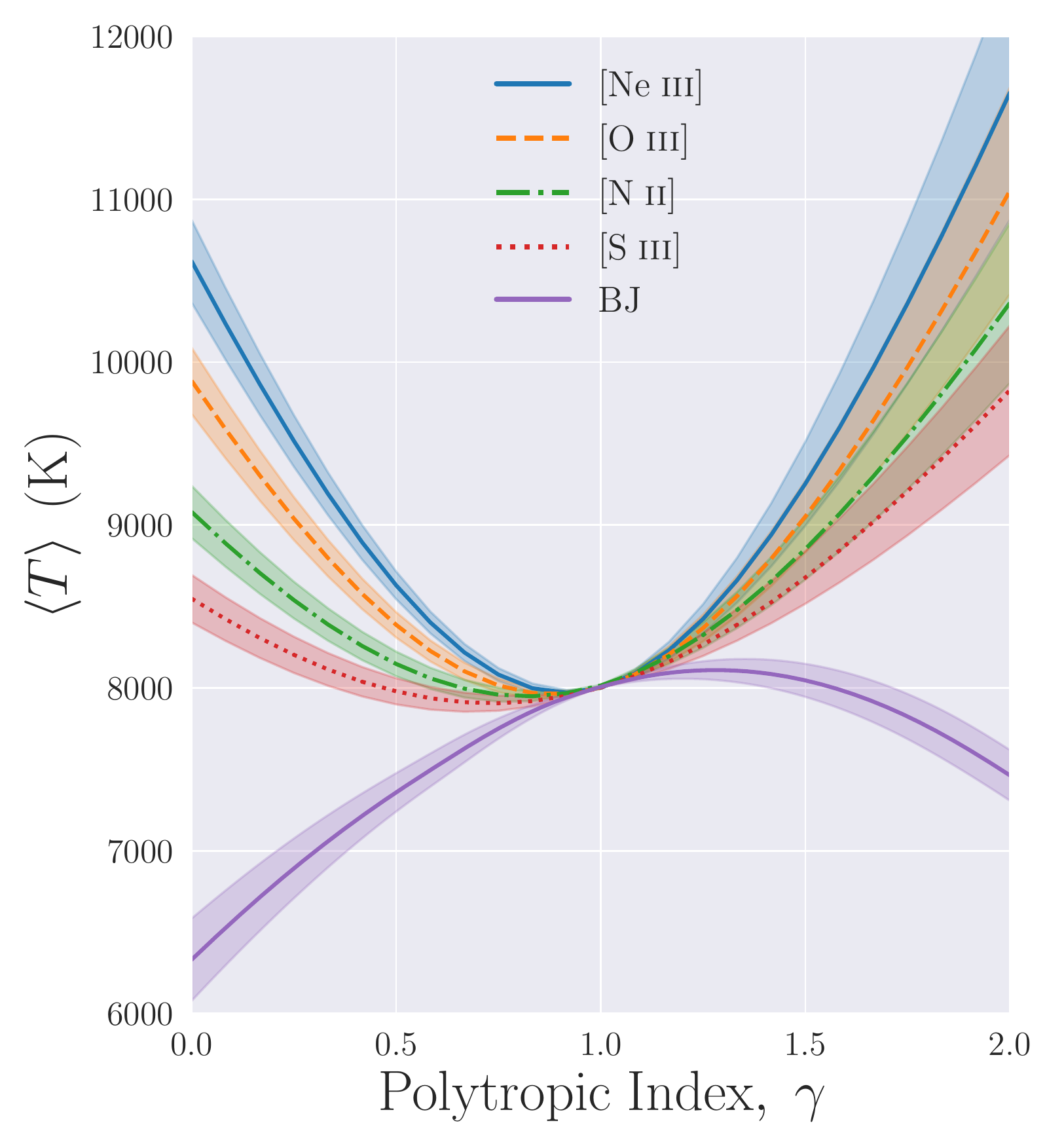}
\caption{Average line of sight temperature estimate $\langle T \rangle$ as a function of the polytropic index $\gamma$ for a density-temperature relationship given by $T \propto n_{e}^{\gamma - 1}$, where the electron densities are randomly drawn from a lognormal distribution with mean $\bar{n}_{e} = 1000 \rm \; cm^{-3}$ and $\sigma = 0.3$. The mean electron temperature was set to $\overline{T} = 8000$ K.}
\label{fig:poly_tem}
\end{figure}

\begin{figure}
\includegraphics[width=\columnwidth]{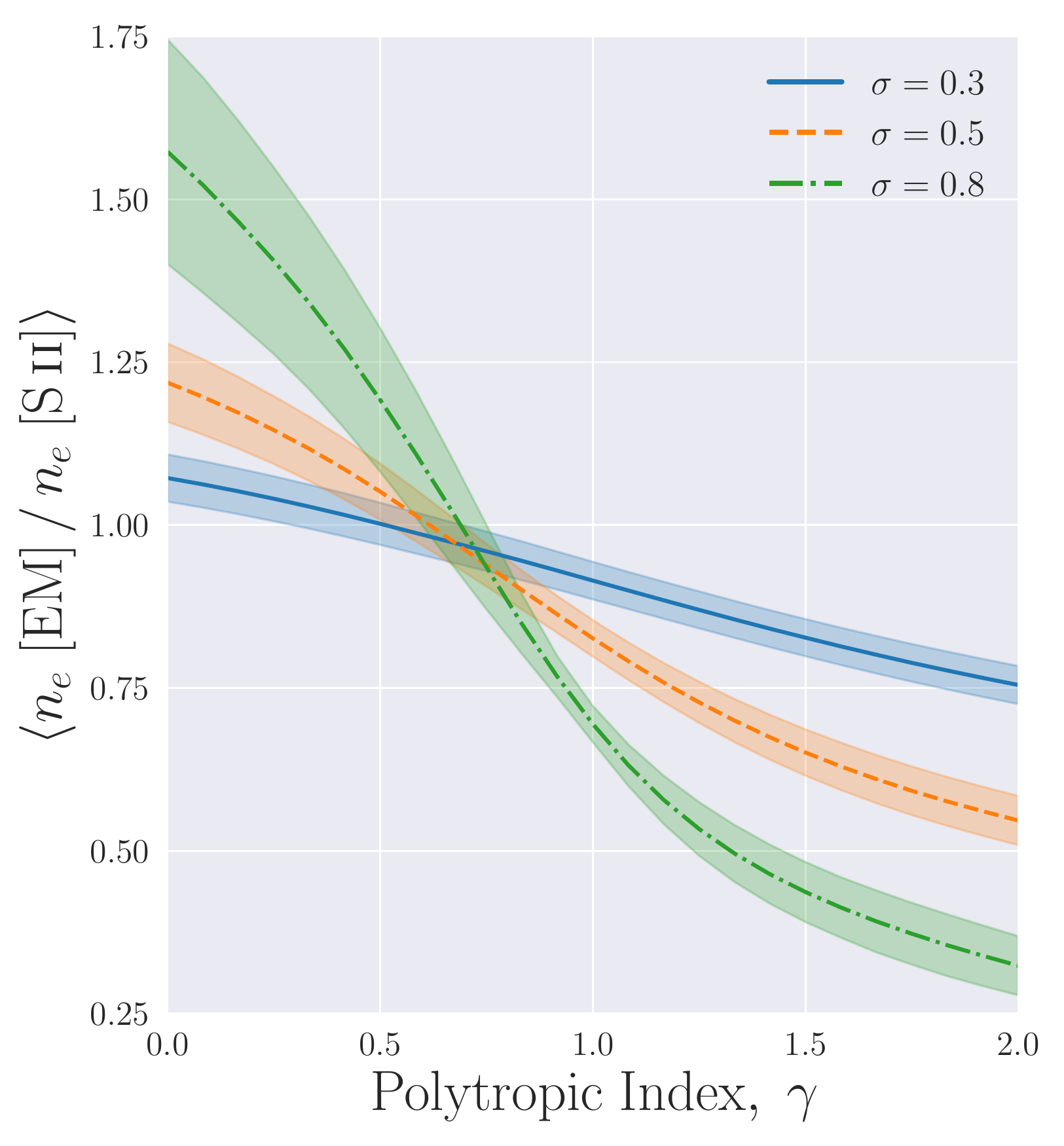}
\caption{Average line of sight ratio of density estimates as a function of the polytropic index $\gamma$ for a density-temperature relationship given by $T \propto n_{e}^{\gamma - 1}$, where the electron densities are drawn from a lognormal distribution with mean $\bar{n}_{e} = 1000 \rm \; cm^{-3}$. The mean electron temperature was set to $\overline{T} = 8000$ K.}
\label{fig:poly_ff}
\end{figure}

\begin{figure*}
\subfloat[{[\ion{O}{iii}] 5007 {\AA}}]{\includegraphics[width=\columnwidth]{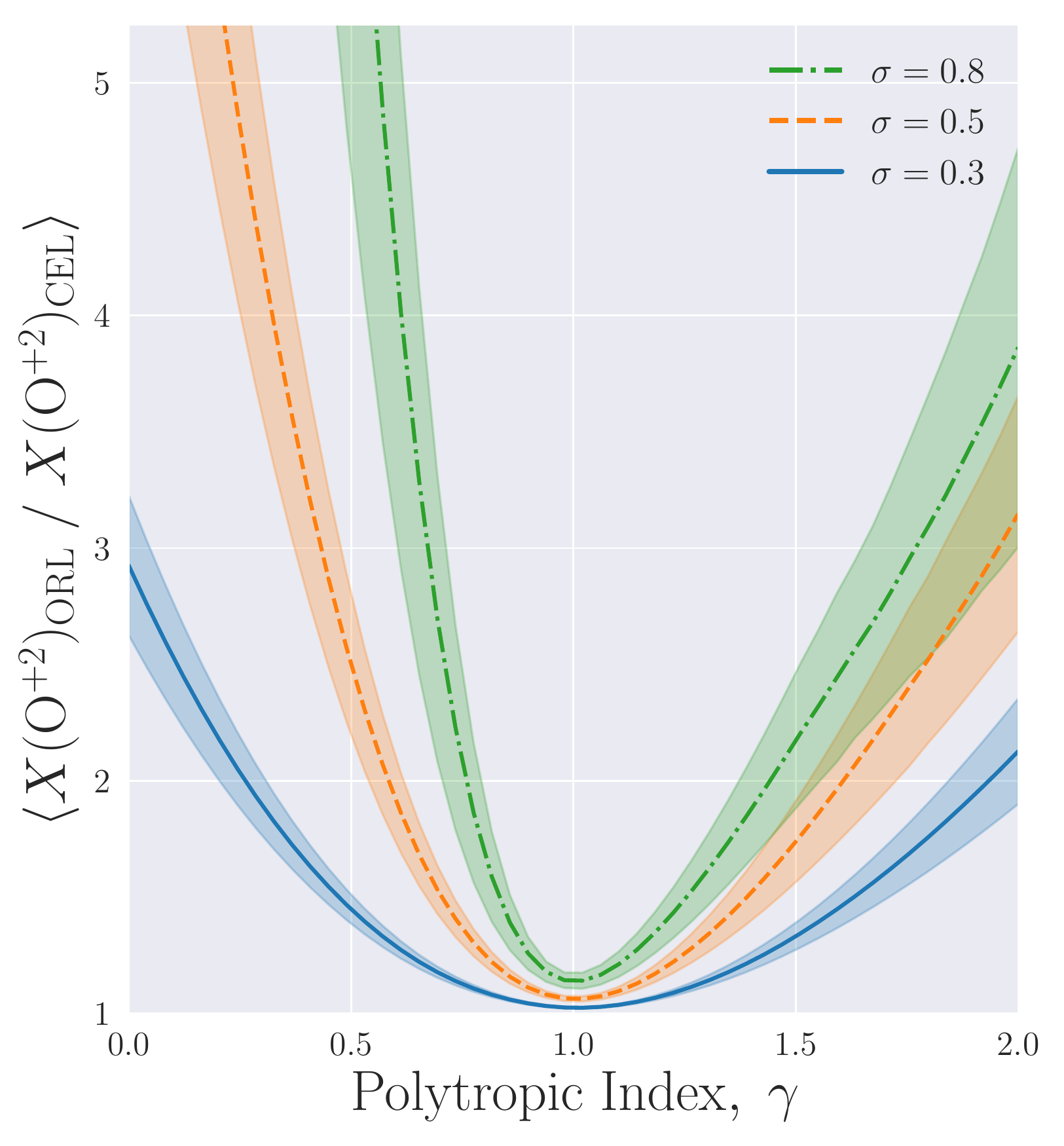}\label{fig:poly_adf_a}}
\subfloat[{[\ion{O}{iii}] 88 $\mu$m}]{\includegraphics[width=\columnwidth]{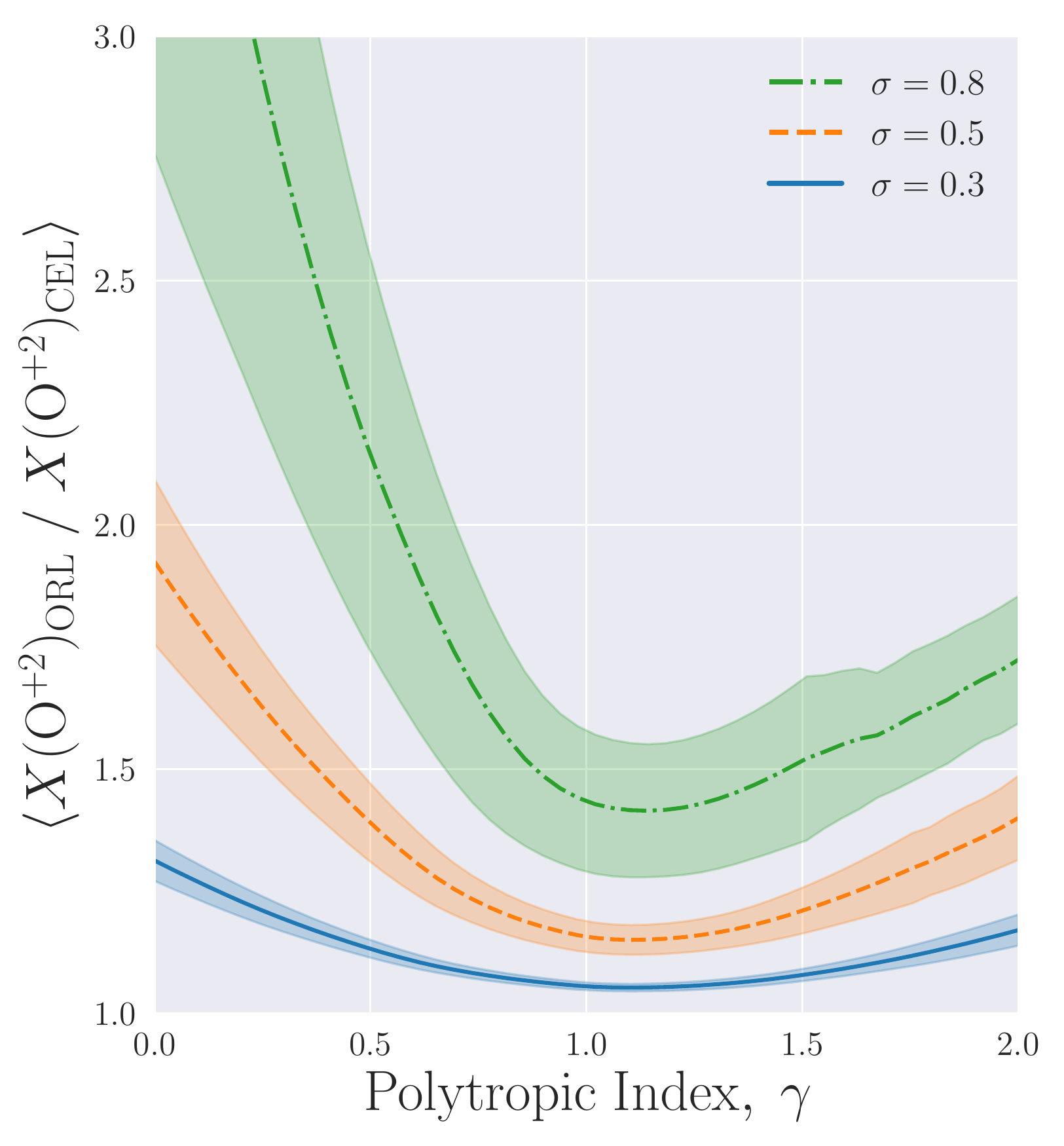}\label{fig:poly_adf_b}}
\caption{Average line of sight ADF value for various polytropic relationships between the density and temperature, with $T \propto n_{e}^{\gamma - 1}$. The mean temperature is set to 8000 K, while the electron densities are drawn from a lognormal distribution with mean $\bar{n}_{e} = 1000 \; \rm cm^{-3}$. The ribbon represents the one standard deviation spread of the observed values.}
\label{fig:poly_adf}
\end{figure*}

Some correlation between density and temperature is to be expected. Denser regions will experience a greater degree of collisional de-excitation, reducing radiative cooling and increasing the temperature \citep{2006agna.book.....O}. Meanwhile, assuming that different regions are in pressure equilibrium leads to the condition $T \propto n_{e}^{-1}$ for the case of an ideal gas. If adiabatic expansion occurs, then $T \propto n_{e}^{\gamma-1}$, where $\gamma = 5/3$ for a monatomic gas.

Since we don't know the equation of state for turbulent fluctuations in photoionised nebulae, we assume a polytropic relationship of the form $T \propto n_{e}^{\gamma - 1}$, where $\gamma$ is allowed to vary over the range $0 \le \gamma \le 2$ to cover a range of conditions, including those mentioned above. The electron densities are drawn from a lognormal distribution with mean $\bar{n}_{e} = 1000 \; \rm cm^{-3}$. The temperature is then calculated for a given $\gamma$ and normalized so that the mean electron temperature is $\overline{T} = 8000$ K.\footnote{Changing the temperature or density doesn't greatly affect the qualitative results that follow.}

Fig. \ref{fig:poly_tem} shows the average line of sight temperature estimate for various diagnostics under the assumption $n_{e} = 1000 \, \mathrm{cm}^{-3}$. When density and temperature are coupled, temperature estimates from CELs exceed those from the Balmer jump. There is a general difference between a positive $(\gamma > 1)$ and negative $(\gamma < 1)$ correlation in that the temperature estimates from different CELs are more overlapping in the former case. When temperatures are positively correlated with density, temperatures derived from the Balmer jump are close to the mean electron temperature.

Fig. \ref{fig:poly_ff} shows the results for the filling factor using the [\ion{S}{ii}] line ratio. The [\ion{S}{iii}] line ratio was used to estimate the temperature along each line of sight for the CELs, while the Balmer jump was used as a proxy for the radio temperature measurement as they give similar estimates \citep{1967ApJ...150..825P}. For $\gamma \lesssim 0.5$, the density inferred from the emission measure exceeds that of the spectroscopic line ratio, which is generally not encountered in real observations. The few instances of filling factors exceeding unity are thought to originate from errors in estimating the size of the nebula \citep{1994A&A...284..248B}. It is worth noting that $\gamma = 0$ corresponds to the physically-reasonable case of pressure-balance fluctuations.

Fig. \ref{fig:poly_adf} shows the ADF results using [\ion{O}{iii}] as the temperature diagnostic for the CELs. Large ADFs can result when temperature and density are negatively correlated for both optical and far-infrared CELs, although the effect is more substantial among optical lines. For $\gamma > 1$, optical CELs can have large ADFs when temperature fluctuations are sufficiently large, while far-infrared CELs only experience a more moderate increase relative to those driven by density fluctuations ($\gamma = 1$). 

Compared to the case with temperature-only fluctuations, we find that a negative correlation between density and temperature increases the ADF of optical CELs whereas a positive correlation lowers the ADF. This is most evident in comparing the $\sigma = 0.3$ lognormal pdf for $\gamma = 0$ and $\gamma = 2$ in Fig. \ref{fig:poly_adf_a} with Fig. \ref{fig:tem_adf_a} for a mean temperature $\overline{T} = 8000 \, \mathrm{K}$. Thus the discrepancies measured by the ADF and filling factor appear to be opposed, with a positive correlation helping to reduce the filling factor and ADF, while a negative correlation increases the ADF and filling factor, eventually causing the latter to exceed unity.

\section{Conclusions}\label{s:conclusion}
We have investigated the degree to which plasma fluctuations in density and/or temperature could account for the long-standing observations of filling factors (corresponding to highly porous plasmas) and abundance discrepancy factors (ADFs) in photoionised nebulae. This has been done by simulating nebulae in which either the density or temperature vary in a stochastic manner, describable by a probability distribution function (pdf). This has been accomplished by dividing the simulated nebulae into cubes with statistically independent densities/temperatures. The size of these cubes are identified with the outer scale of the turbulence.

With these simulated nebulae we have calculated observable quantities such as the emission measure and the intensities of emission lines associated with density- and/or temperature-sensitive line ratios. These quantities have then been compared to simulate astronomical measurements. The results from this investigation are as follows:

\begin{enumerate}
	\item Density fluctuations with realistic pdfs can reproduce the observational signature conventionally interpreted as evidence for filling factors $f < 1$. Density pdfs with pronounced tails extending to large densities (e.g. lognormal, Pareto) can reproduce values of $f < 0.1$, as indicated in many astronomical observations. Thus it is not necessary to invoke the extreme `dual delta function' model suggested by \cite{1959ApJ...129...26O}.
	\item The discrepancy between the density inferred from the emission measure and that from spectroscopic line ratios also depends on the value of the mean density relative to the critical density of the emission lines. When the true mean density is much less than the critical density, the discrepancy is larger, and a smaller filling factor is inferred. As such, we predict that line ratios with smaller critical densities (e.g. [\ion{O}{ii}], [\ion{S}{ii}]) should produce larger filling factors compared to lines with greater critical densities (e.g. [\ion{Ar}{iv}], [\ion{Cl}{iii}]). Deviations from this would then point to additional factors taking place, such as chemical inhomogeneities.
	\item Density fluctuations with realistic pdfs can also reproduce the observational signature of ADFs when using far-infrared CELs, but only up to around a factor of five or so. Since the density-dependence of the emissivity of ORLs better traces the emissivity of the hydrogen recombination lines by which the hydrogen abundance is determined, the abundance estimates of ORLs are generally more accurate than those from CELs. As the mean density of the nebula increases, these discrepancies become more pronounced as the denser regions disproportionately dominate the light from ORLs compared to CELs.
	\item Nebulae with electron temperature (but not density) fluctuations yield inferred temperatures higher than the true mean temperature when using CELs. The bias depends on the set of temperature-sensitive transitions used (e.g. [\ion{Ne}{iii}] vs. [\ion{S}{iii}]) and the mean temperature. This result suggests that a comparison of temperature inferred from two or more sets of temperature diagnostics, and a comparison with ion temperature measurements, could provide a diagnostic for temperature fluctuations.
	\item Temperature fluctuations are also able to produce the observational signature of ADFs. This becomes most pronounced for optical CELs and at lower mean temperatures. Far-infrared CELs can show moderate ADFs, but only up to around $ADF(\mathrm{O}^{+2}) \lesssim 1.5$ for standard nebular temperatures $(\overline{T} > 5000 \; \mathrm{K})$.
	\item In the case of density and temperature fluctuations being simultaneously present, the observational consequences depend on the form of the pdf and the functional form of the density-temperature relationship. The density-temperature relationship is parametrized by a polytropic index $\gamma$ defined as $\delta T \propto \delta n_{e}^{\gamma - 1}$. An interesting and potentially significant result is that for the seemingly plausible case of $\gamma = 0$ (corresponding to isobaric fluctuations) the density inferred from the emission measure exceeds that inferred from density-dependent line ratios, in contradiction to most observations. Fluctuations where $\delta n_{e}$ and $\delta T$ are positively correlated yield filling factors of the sort observed. This result raises the interesting possibility that real observations can yield information or constraints on the thermodynamics of small scale fluctuations in \ion{H}{ii} regions and PNe.
	\item A polytropic relationship between density and temperature leads to biases in the inferred temperature using different diagnostics. A positive correlation $(\gamma > 1)$ leads to a greater overlap in inferred temperature using various CELs, while the Balmer jump tends to provide a good estimate of the true mean temperature. A negative correlation $(\gamma < 1)$, meanwhile, tends to result in a greater separation between the temperatures inferred from various CELs.
	\item A polytropic relationship between density and temperature can produce large ADFs when the density and temperature are negatively correlated. Such a case, however, appears to be ruled out by observations of the filling factor $f < 1$. If density and temperature are positively correlated, large ADFs can also result if $\gamma$ is sufficiently large when using optical CELs. In such situations, far-infrared CELs may produce lower ADFs compared to their optical counterparts as they are mainly driven by density fluctuations. Such a scenario might take place in \ion{H}{ii} regions, where the abundances inferred from far-infrared CELs exceed those from optical CELs in the Orion Nebula \citep{2016arXiv161203633E}.
\end{enumerate}

In our simulations we have assumed that the ion density is proportional to the electron density, and thus have not taken the distribution of ionization states into consideration. This could be an area of future improvement.\footnote{Our code is available at \href{https://github.com/bbergerud/Nebularium}{github.com/bbergerud/Nebularium}}


\section*{Acknowledgements}
We thank the referee of this paper for a collegial and very helpful review.



\bibliographystyle{mnras}
\bibliography{report} 





\bsp	
\label{lastpage}
\end{document}